\documentclass[%
 reprint,
superscriptaddress,
showpacs,
showkeys,
 amsmath,amssymb,
 aps,
floatfix,
]{revtex4-1}
\usepackage{graphicx}
\usepackage{mathrsfs}
\usepackage{amsmath,bbold}
\usepackage{caption}
\usepackage{subcaption}
\usepackage{tabularx,ragged2e,booktabs,caption}
\usepackage{latexsym}
\usepackage{color}
\usepackage[colorinlistoftodos]{todonotes}
\usepackage{subfig}
\usepackage{datetime}
\usepackage{enumerate}
\usepackage{chngcntr}

\counterwithin*{paragraph}{section}

\DeclareMathOperator{\Tr}{Tr}
\begin{document}
\title{Temporal stability of network partitions}
\author{Giovanni Petri}
\affiliation{ISI Foundation, via Alassio 11/c, 10126 Turin, Italy}
\author{Paul Expert}  
\affiliation{Centre for Neuroimaging Sciences, Institute of Psychiatry, De Crespigny Park, King's College London, London SE5 8AF, UK}
\begin{abstract}
We present a method to find the best temporal partition at any time-scale and rank the relevance of partitions found at different time-scales. This method is based on random walkers coevolving with the network and as such constitutes a generalization of partition stability to the case of temporal networks. We show that, when applied to a toy model and real datasets, temporal stability uncovers structures that are persistent over meaningful time-scales as well as important isolated events, making it an effective tool to study both abrupt changes and gradual evolution of a network mesoscopic structures.
\end{abstract}
\date{\today,\currenttime}
 %89.75.Kd: patterns, 89.75.Hc: Network and genealogical trees, 89.75.-k: complex systems, 89.75.Fb: Structures and organization in complex systems, 05.40.Fb: Random walks and Lévy flights
\maketitle

\section{Intro}
Identifying mesoscopic structures and their relation to the function of a system in biological, social and infrastructural networks is one of the main challenges for complex networks analysis \cite{Fortunato:2010vi}. 
Until recently, most approaches focused on static network representations, although in truth most systems of interests are inherently dynamical \cite{Holme:2012jo}. 
Recent theoretical progresses and the availability of data inspired a few innovative methods, which mostly revolve around unfolded static representations of a temporal dynamics \cite{Mucha:2010bz,Ronhovde}, constraints on the community structure of consecutive graph snapshots \cite{Kawadia:2012ev,Chen:2013uw} or on global approaches
% (e.g. non-negative matrix decompositions of network tensor representations
\cite{Kolda:2009dh,Gauvin:2013waa,Lin,Lancipone2011}.\\
\indent In this Article, we take a different route and tackle the problem of finding and characterising the relevance of community structures at different time-scales by directly incorporating the time-dependence in the method. \\
\indent Inspired by the notion of {\it stability} \cite{Delvenne:2010vx}, we propose a related measure, {\it temporal stability}, which naturally embeds the time-dependence and order of interactions between the constituents of the system. Temporal stability allows not only to compare the goodness of partitions over specific time scales, as its static counterpart, but also to {\it find} the best partition at any time and over any time scale.
In the following, we briefly review the main ingredients of static stability and introduce their natural extensions to temporal networks. We then present a benchmark model as a proof of principle, and then analyse two real-world datasets, finding pertinent mesoscopic structures at different time-scales.

\section{Temporal stability}
Like the map equation \cite{Rosvall:2010cy,Rosvall:2009wj}, stability exploits the properties of the stationary distribution random walkers exploring a static network and of long persistent flows on a network.
While the map equation relies on finding the most compressed description of a random walker trajectory in terms of its asymptotic distribution, the intuition behind stability is that walkers exploring the network will tend to stay longer in a well defined cluster before escaping to explore the rest of the network. 
The object of interest is thus the auto-covariance matrix of an unbiased random walk on a network $G$ for a given partition $\mathscr{H}$, i.e. the higher the autocorrelation, the better the description of a system in terms of modules by $\mathscr{H}$. 
After $\tau$ Markov time-steps of exploration of the network by the random walkers, it can be compactly written as:
\begin{equation} \label{autocorr}
R_\tau  = \mathbf{H}^T (\Pi \mathbf{M}_G^\tau - \pi^T \pi ) \mathbf{H}
\end{equation}
where $\mathbf{H}$ is the partition matrix assigning nodes to communities, $\mathbf{M_G}$ the transition matrix of the random walk on $G$, $\pi$ its stationary distribution and $\Pi = diag(\pi)$ \cite{Delvenne:2010vx}. The $ \pi^T \pi$ term can be interpreted as a null model that represents the asymptotic modular structure against which the structure unveiled by the random walkers' exploration of the network is tested.
The stability of partition $\mathscr{H}$ at Markov time $\tau$ is then defined by:
\begin{equation}\label{stability}
r_{\mathscr{H}} = \min_{0\leq s < \tau} \Tr{R_s}.
\end{equation}
The magnitude of the trace of the autocovariance matrix represents the extent to which walkers are confined within the clusters defined by $\mathscr{H}$. The minimum over the Markov time during which the walkers were allowed to move ensures that the measure conservative.
The value of $\tau$ at which a given partition becomes optimal conveys information about which topological scales of the network are best described by the partition considered. Moreover, the interval over which a partition is optimal is related to the importance of that specific scale across the hierarchy of scales present in the network. \\ 
%%%%%%%%%%%%%%%
\noindent Extending this measure to temporal networks requires generalizing its ingredients: the partition $\mathscr{H}$, the transition matrix $M_{G}$ and the asymptotic walker distribution $\pi$. 
\paragraph{Temporal partition.}
Let us define a discrete temporal network $G\{ t \}$ as a time-ordered collection of graph snaphots,  $G\{t\} = \{G_0, G_1, \ldots, G_T\}$, represented by their adjacency matrices $\{\mathbf{A}_0,\mathbf{A}_1, \ldots, ,\mathbf{A}_T\}$.
The static partition matrix is naturally extended to the temporal case by allowing it to be time-dependent, $\mathscr{H} \to \mathscr{H}\{t\} = \left \{ \mathbf{H}(t) | t=0,1,\ldots T \right\}$, with $T$ being the number of slices in the temporal dataset. At this point it is worth noting that $\mathscr{H}\{t\}$, like $G\{ t \}$, does not need to change at every time step.\\
\paragraph{Transition matrix.}
The transition matrix $\mathbf{M}$ will in general change between time steps and therefore one does not simply iterate it. We define $\mathbf{M}_t$ the single-snapshot transition matrix relative to $G_t$ as $\mathbf{M}_t = \lim_{\epsilon\to 0} (\mathbf{D_t})^{-1} (\epsilon \mathbf{\mathbb{1}} + \mathbf{A}_t)$, with $\mathbf{D}_t =  (\epsilon \mathbf{\mathbb{1}} + \mathbf{A}_t) \cdot \mathit{1}$, where $\mathit{1}$ is the constant vector with unit components. 
The $\epsilon$-limit is equivalent to including a self-loop of vanishing weight and is required to ensure that the transition matrix is well-defined for nodes that are disconnected at time $t$  
\footnote{While this is not necessary in the case of static stability, because disconnected components can be analysed separately, it is extremely important for random walkers moving on an evolving network, where such components can drastically change from one slice to the next.}.
Using the matrices $M_t$ we can define a time-ordered product $\mathcal{M}_G(t,\tau) $ which represents the transition matrix for the evolution of the random walker across the changing network between $t$ and $t+\tau$:  
%%%
\begin{equation}\label{temporal_autocorr}
\mathcal{M}_G(t,\tau) =  \mathcal{T} \{ \mathbf{M}_t \ldots \mathbf{M}_{t+\tau} \} =  \prod_{s=t}^{t+\tau} \mathbf{M}_{s} =\mathbf{M}_{t+\tau}\cdot\ldots\cdot\mathbf{M}_{t}
\end{equation}
where we impose right multiplication to respect the arrow of time.\\
\paragraph{Stationary walker distribution.} The last element we need is the stationary walker distribution on a time-varying network. We will note this distribution with $\omega$. Different types of random walks can be devised, depending on the model for the dynamics of the network, therefore $\omega$, is not unique; different dynamics preserving different statistical features of the system. 
Here we consider the case where the time-evolving connectivity of each node is known. 
The activity-driven model, introduced by Perra et al. \cite{Perra:2012fi} is particularly adapted to such systems.
It provides a null model akin to a temporal configuration model where the nodes’ temporal activities play the role of the nodes’ degrees. We only introduce here the main concepts needed for our purpose, but a full description of the model is given in the Appendix.
Importantly, the stationary distribution for walkers coevolving with the network \cite{Perra:2012ek} is analytically amenable and provides a natural null model for temporal stability.
The stationary walker distribution for a node with activity $a$ is given by:
\begin{equation}\label{activity_walker_distribution}
\omega_a = \frac{amw + \phi}{ a + m\langle a \rangle}
\end{equation}
where $w$ is the average density of walkers on a node, $\phi$ a scalar that can be obtained numerically in closed form and $\langle a \rangle$ the average activity. We used $m=2$ and the activities were computed such that the temporally averaged degree is conserved \cite{Perra:2012fi}.
Without loss of generality, we can set $w=1/N$, $N$ being the number of nodes in the network, and use Eq.~(\ref{activity_walker_distribution}) for the stationary walker distribution.\\% (see Appendix for more details). \\ % in Eq.~(\ref{temporal_autocorr}).\\

We are now in a position where we can define the {\it temporal stability} for a partition $H\{t \}$ of $G\{ t \}$ at time-scale $\tau$:
\begin{equation} \label{temporal_stability}
r_{\tau, \mathscr{H}\{t\}}  =  \left \langle  \Tr R_{t',\tau}  \left [G\{t\},  \mathscr{H}\{t\} \right ] \right \rangle_{t'},
\end{equation}
with 
\begin{equation}\label{temporal_autocorr}
R_{t', \tau}  \left [G\{t\},  \mathscr{H}\{t\} \right ]  =  \mathbf{H}^T(t') \left [\mathbf{\Omega}\mathcal{M}_G(t',\tau) - \omega^T \omega \right ]  \mathbf{H}(t'),
\end{equation}
where $\mathbf{\Omega} = diag (\omega)$ and the average over $t'$ in Eq.~(\ref{temporal_stability}) is taken over $[0,T] $ and plays a similar role as the minimum over $\tau$ in the static stability. The trace is taken inside the temporal average to allow for partitions of different sizes at different times. Temporal stability is naturally interpreted as the average stability obtained over all windows of size $\tau$ for a given temporal partition $\mathscr{H}\{ t \}$.\\
\indent In addition to providing a natural measure to evaluate the relevance of partitions over different time-scales $\tau$, temporal stability characterises the partition $\mathscr{H}^{opt}[t,\tau]$ with the highest stability for every pair ($t,\tau$). 
By linearity, the average over $t$ can be unfolded, leaving the expression:
\begin{equation}\label{Optimal}
\mathbf{B}(t,\tau) = \left [\mathbf{\Omega}\mathcal{M}_G(t,\tau) - \omega^T \omega \right].
\end{equation}
The partition $\mathscr{H}^{opt}[t,\tau]$ optimising \eqref{Optimal} is the same as the one optimising temporal stability at time $t$ over the time-scale $\tau$ and it can be found by considering Eq. (\ref{Optimal}) as a modularity optimisation problem and solving it with any standard modularity algorithm (e.g.  the Louvain method \cite{Blondel:2008do}). 
%This approach is similar to the one described in \cite{lambiotte-2008}  for static stability.
In the Appendix we give  a detailed description of the workflow and provide a link to our code.
%Note that the optimal temporal partition $\mathscr{H}^{opt}[t,\tau]$ contains information not only on the partition at a given $t$ but also on its relevance for every other possible $\tau$, as with any partition.\\
\section{Two blocks toy model}
\begin{figure}
\centering
\includegraphics[width=0.35\textwidth]{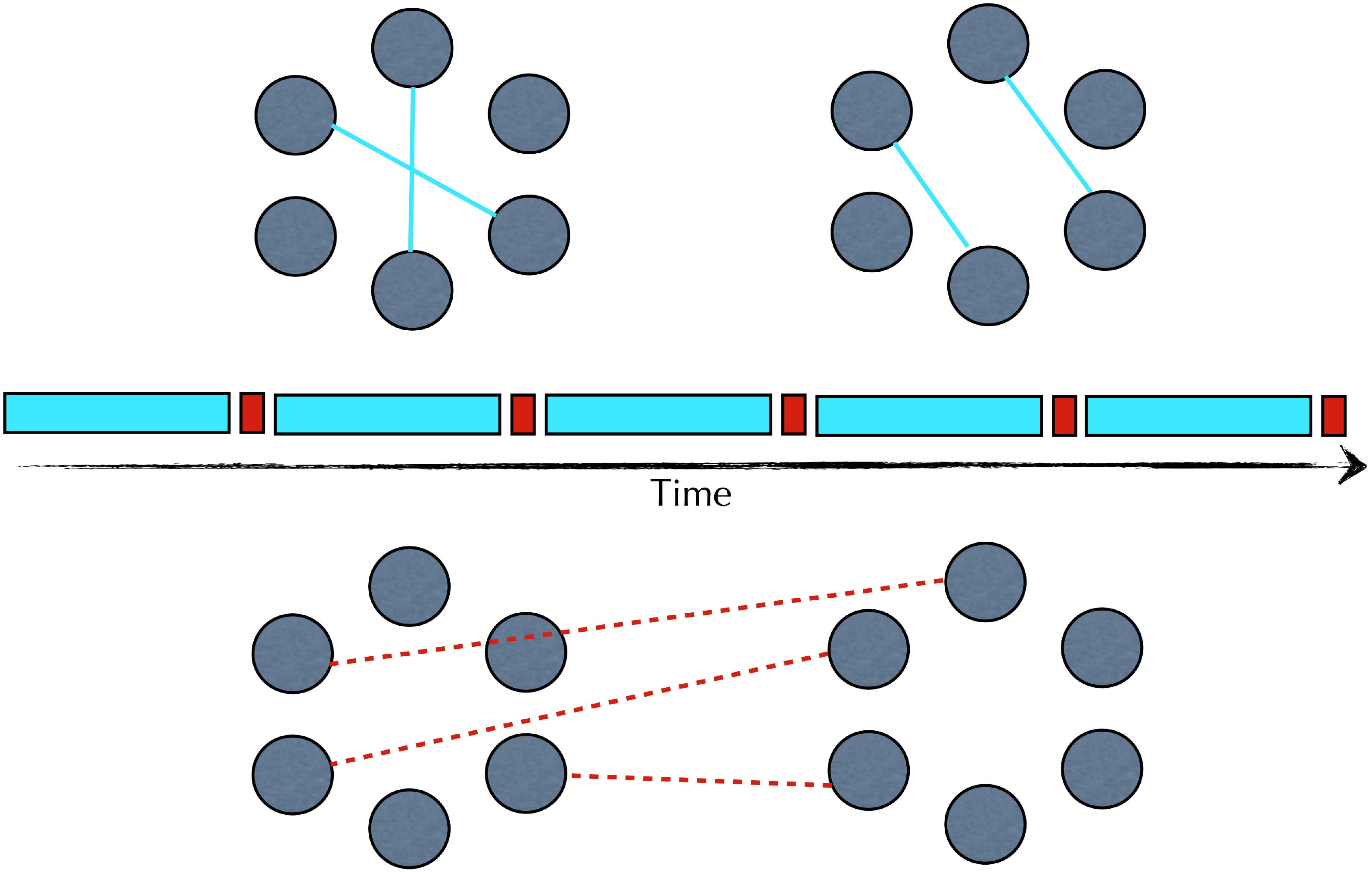}
\caption{{\bf Two blocks model} The two blocks of nodes interact following simple alternating rules: during a time window of $T_{in}$ timesteps ({\it blue (colour)/long (grey) intervals}), nodes can create connections only to nodes within their block (top) with probability $p_{in}$; then during a time window $T_{out}$ ({\it red (colour)/ short (grey) intervals}), they are allowed to interact only with nodes belonging to the other block ({\it bottom}) with probability $p_{out}$. %For each snapshot, connectivity patterns are obtained as Erd\"os-Renyi networks. 
The linking probability for connections within  $p_{in}$ and between blocks $p_{out}$ are chosen such as to make the time aggregated network a single uniform hairball, hiding the temporal nature of the division in two blocks.} \label{fig::two_block_sketch}
\end{figure}

\begin{figure}[h]
\includegraphics[width=0.5\textwidth]{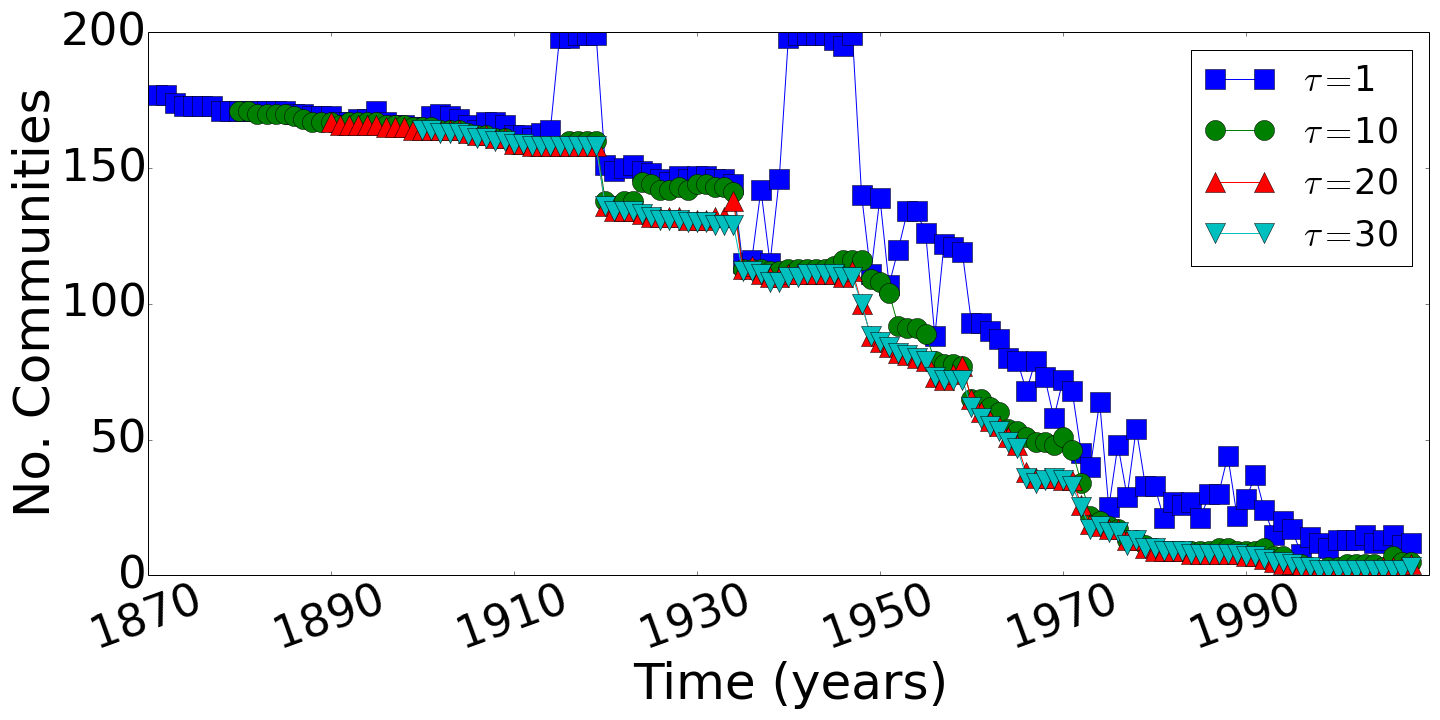}
\caption{Comparison of the number of communities in optimal partitions for different values of $\tau$ in the International Trade Network.} \label{fig::cow_comparison_tau_partitions}
\end{figure}

A simple example is useful to illustrate the working of temporal stability and the type of structure it can unveil.
Following the model with activity-correlated link classes defined in \cite{Gauvin:2013uy}, we consider a toy model network consisting of two blocks of $N$ and $N'$ nodes that follow a simple cycle of temporal interactions repeated $M$ times.  
Each cycle consists of two interaction windows with a total time period of $T=T_{in}+T_{out}$, see Fig.(\ref{fig::two_block_sketch}) for an illustration. During the first $T_{in}$ time-steps, the nodes only interact within their block with a probability $p_{in}$ per time-step; in the remaining $T_{out}$ time-steps, the nodes interact exclusively with nodes from the other block with probability $p_{out}$. 
Setting $p_{out}=\frac{T_{in}p_{in}}{T_{out}}\frac{(N(N-1)+N'(N'-1)}{2NN'}$, guarantees that the density of links within and between blocks is the same. 
This makes it impossible to distinguish the two communities at the time-aggregated level.\\
\indent We then compare the stability of various partitions obtained by aggregating the temporal network over time windows of different size $\Delta=10,20,80$ time-steps and the bipartition into the two blocks (Fig. \ref{fig::sociopatterns_stability}a  {\it top}).
%%%%%%%
%%%%
These partitions represent progressively coarser temporal summaries of the network and we refer to them as $\Delta$-partitions. They show how aggregating temporal networks wipes out relevant dynamics.
With the simulation set-up we used, a crucial value for time is $t=20$
%(Fig. \ref{fig::two_block_sketch})
as it represents the interval of time over which the two blocks are well defined. 
This is clearly shown by the value of $r_{\tau}$ for the bipartition which steadily decreases towards zero for $\tau\geq T_{in}$.
This illustrates well how temporal stability behaves: the bipartition is a reasonably good approximation over time-scales shorter than the mixing cycle, while  for longer time-scales, the bipartition description is lost in the noise. 
%The effect and meaning of the time-scale $\tau$ is discussed in more details in the next section.
The parameter $\tau$ can then be used to select a time-scale over which to compare the relevance of partitions. 
A partition that is relevant over a short time-scale captures essential micro-dynamics, while at longer time-scales other global mechanisms prevail and the micro-dynamics can be considered as noise.
To illustrate this phenomenon, we plot in  in Fig. \ref{fig::cow_comparison_tau_partitions} the simplest quantity, the number of communities in a partition, for different values of $\tau$ in the trade network dataset (which we fully introduce in the next section).
The shortest time-scale $\tau=1$ yr  clearly identifies the World Wars and other events that are precisely located in time. 
As soon as one switches to a time-scale longer than those events, e.g. $\tau=10$yr, these effects disappear and the number of communities displays a much smoother behaviour over time and captures a different type of dynamics, namely the densification of links in the network. 
This effect is even stronger if one considers longer time scales $\tau=20,30$yr.\\
\indent Thus, in our toy model, $\Delta$-partitions with $\Delta<T$  perform well for small $\tau<T_{in}$, as the aggregation window allows for a finer temporal sampling of the network's evolution. Their stability values then continuously decay for growing $\tau\geq T_{in}$ and eventually tend to 0. Moreover, the $\Delta$-partitions with $\Delta >T$ perform better for $\tau\geq T_{in}$ as the mixing windows are incorporated in the aggregated networks.
Figure \ref{fig::sociopatterns_stability}a also displays the temporal stability for the optimal partition, found with Eq. (\ref{Optimal}), and which outscores all other partitions for all $\tau$.\\
%%%%%%
\begin{figure}[h]
\begin{subfigure}[t]{0.4\textwidth}
\includegraphics[width=\textwidth]{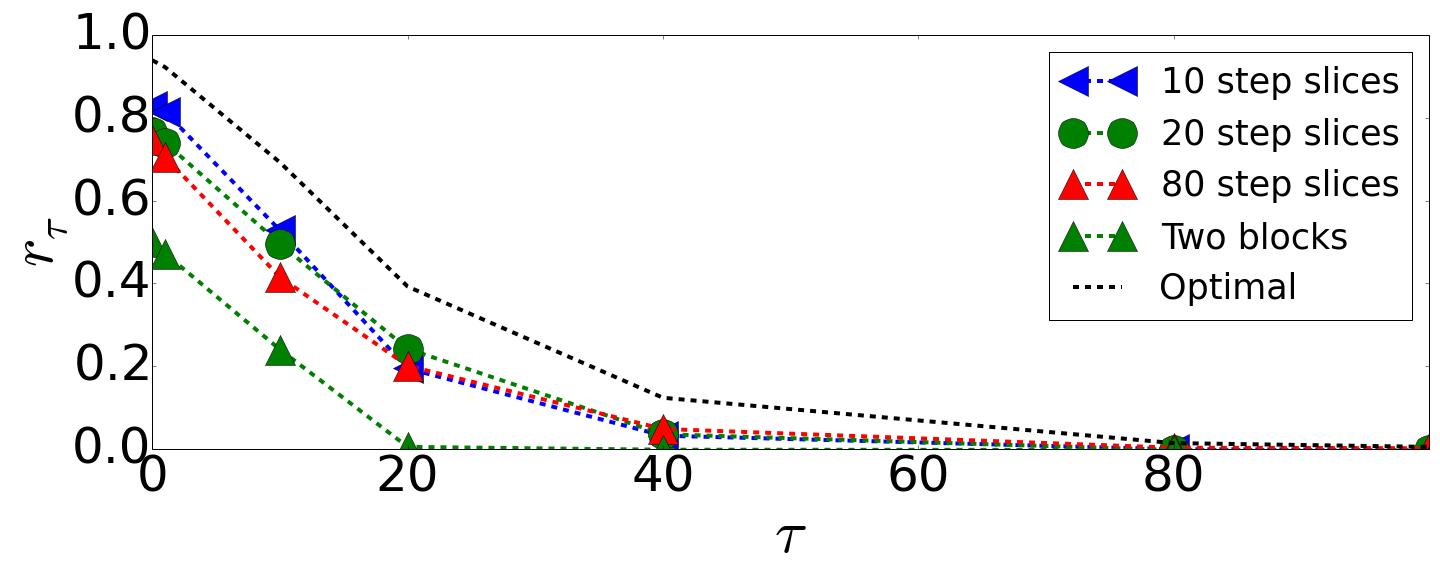}
\includegraphics[width=\textwidth]{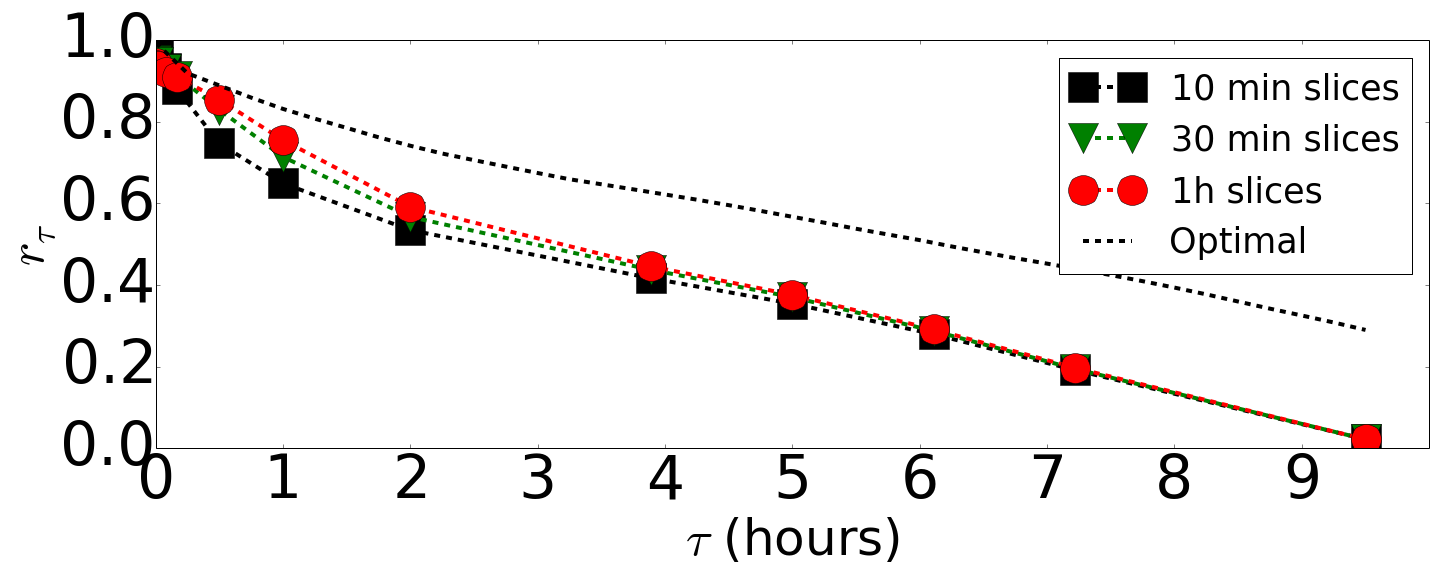}
%\caption{Temporal stability $r_{\tau}$ curves for the two blocks model ({\it top}) and Sociopatterns ({\it bottom}) for different $Delta$-partitions and the optimal partitions.}
\end{subfigure}
\begin{subfigure}[t]{0.4\textwidth}
\centerline{\includegraphics[width=1.3\textwidth]{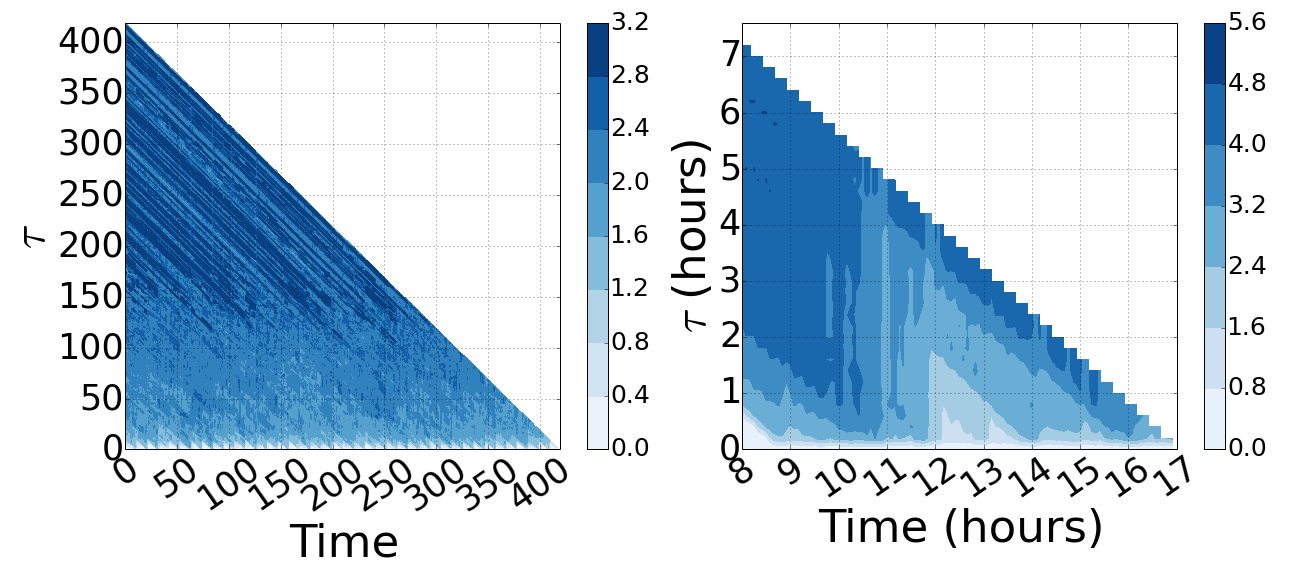}}
%\caption{Variation of information $v(t,0,\tau)$ for the two blocks model ({\it left}) and Sociopatterns ({\it right}).}
\end{subfigure}
\caption{{\bf Comparison of two blocks model and Sociopatterns data} Panel (a) shows the temporal stability curves $r_\tau$ for the two blocks model ({\it top}) and Sociopatterns ({\it bottom}) for a selection of $\Delta$-partitions against the best temporal clustering obtained from Eq. (\ref{Optimal}). Moreover, temporal stability identifies the periodic nature of the two datasets with the recurring sawtooth pattern evident in the variation from the information $v(t,0,\tau)$ in (b). The two blocks model is on the ({\it left}) and Sociopatterns on the ({\it right}). The variation of information is computed from the optimal partitions.
The parameters for the two-block model are $N=N'=10$, $T_{in} = 20$, $p_{in}=0.01$, $T_{out}=1$ and $p_{out} =0.18$.}
\label{fig::sociopatterns_stability}
\end{figure}
%%%%%%%%%%%%%%%%%%%%%%%%%
The {\it variation of information} $v(t,\tau,\tau')$ between $\mathscr{H}^{opt}[t,\tau]$ and $\mathscr{H}^{opt}[t,\tau']$ is a useful tool to detect structural 
changes between clusterings \cite{Fortunato:2010vi,Meila:2003gp}. 
Consider two partitions of a network $\mathscr{X}$ and $\mathscr{Y}$ in $k$ and $l$ communities respectively.
Associate to each community in the partition a probability proportional to its cardinality, for example to $X_i \in \mathscr{X} $ associate $p^X_i = |X_i|/N$, where $N$ is the number of nodes in the network under consideration.
Finally, $v(t,\tau,\tau')$ is defined as:
\begin{eqnarray}\label{VI_def}
v(t,\tau,\tau') = H(\mathscr{H}^{opt}[t,\tau]) + H(\mathscr{H}^{opt}[t,\tau']) \nonumber \\
- 2I(\mathscr{H}^{opt}[t,\tau] ,  \mathscr{H}^{opt}[t,\tau'],
\end{eqnarray}
where $H$ is the Shannon entropy and $I$ the mutual information defined on the probabilities introduced above. \\
Two special cases are particularly enlightening regarding the evolution of the optimal partition in time: $v(t,0,\tau)$, which informs about the time-scale over which structures persist from a time $t$ (see Fig. \ref{fig::sociopatterns_stability}b left and \ref{fig::cow_stability}b left) and $v(t,\tau,\tau+1)$ which gives the instantaneous structural changes between two consecutive time steps $t+\tau$ and $t+\tau+1$ (see Fig. \ref{fig::cow_stability}b right), starting at time $t$. In the first case, key informations about stability are read on the ordinate axis and in the second case, sudden changes can be read on the abscissa axis.
In the two blocks model, information about the state at time $t$ is almost completely lost after two cycles, although this depends on when the initial time $t$ is in the cycle: at the beginning of a cycle a walker is constrained within a block and thus the community structure is stable for a longer time; the inter-block interaction then comes as a shock for the community structure, producing the saw-like pattern evident in Fig. (\ref{fig::sociopatterns_stability}b left).
Since the toy model has no additional structure on top of the periodicity, we expect $v(t,\tau,\tau+1)$ to show only noise due to fluctuations in the connectivity within blocks.\\
%%%%%%%% 	sociopatterns %%%%%%%%%%
\section{Real datasets}
\begin{figure}
\centering
\begin{subfigure}[t]{0.4\textwidth}
\includegraphics[width=\textwidth]{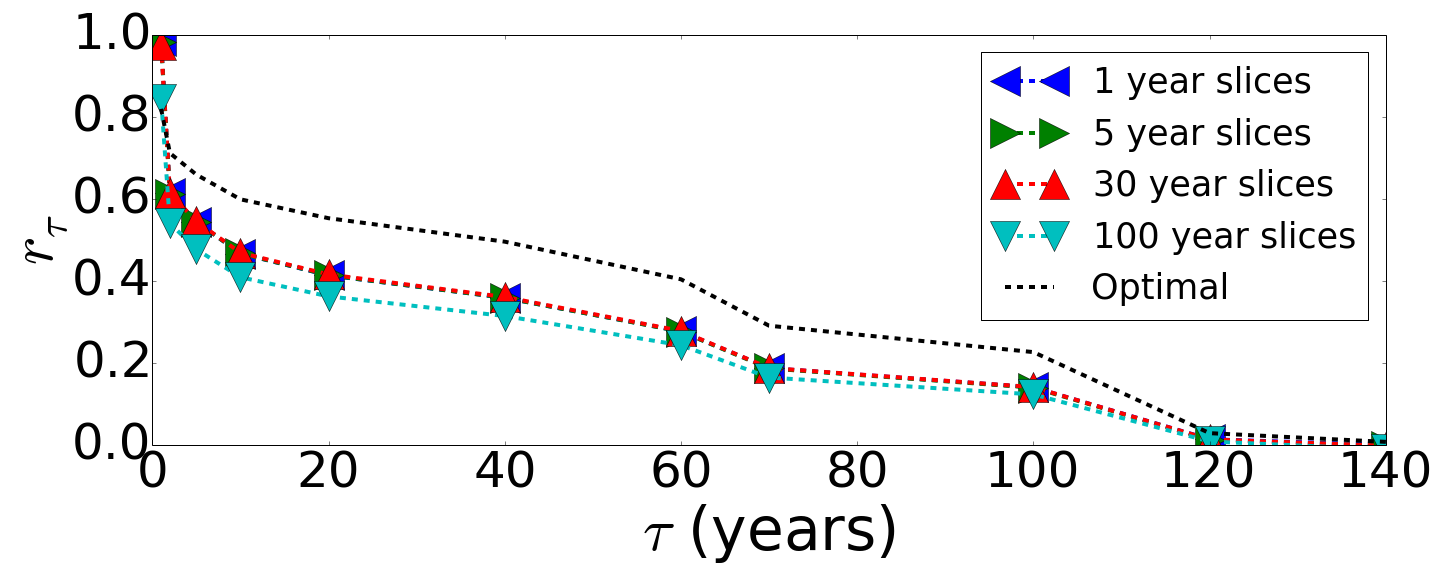}
%\caption{Temporal stability curves for the international trade network for different $\Delta$-partitions and the optimal partitions}
\end{subfigure}
\begin{subfigure}[t]{0.4\textwidth}
\centerline{\includegraphics[width=1.3\textwidth]{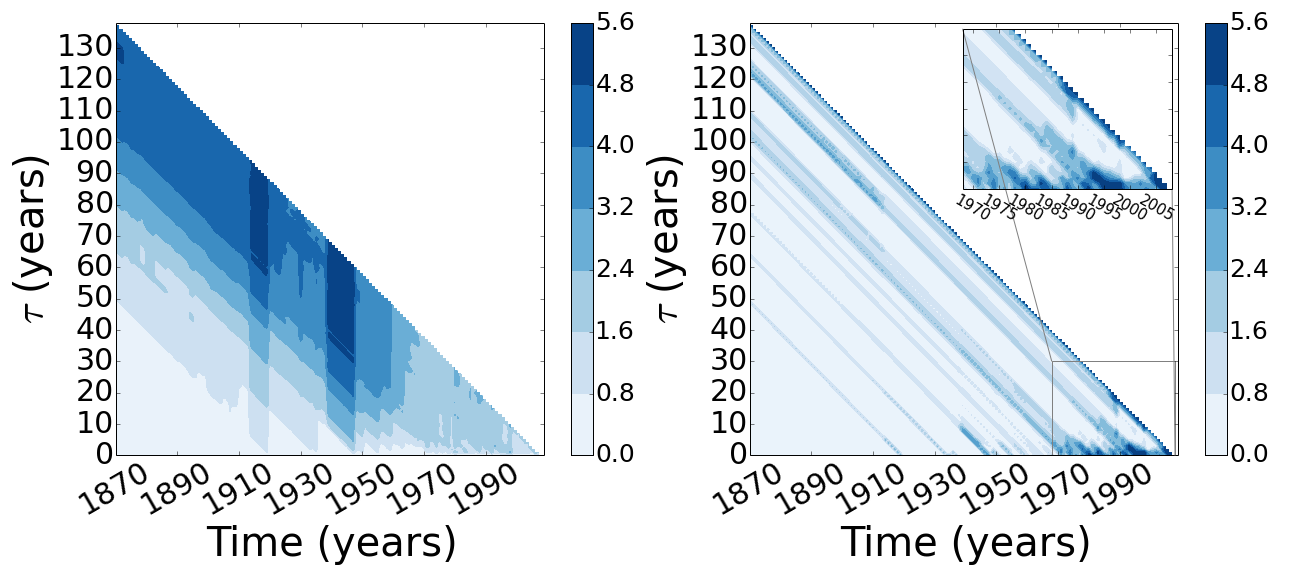}}
%\caption{Variation of information $v(t,0,\tau)$ ({\it left}) and $v(t,\tau,\tau+1)$ ({\it right}) for the international trade network.}
\end{subfigure}
\caption{{\bf Financial trade network}  In (a), we show the temporal stability curves $r_{\tau}$ for different $\Delta$-partitions. The optimal temporal clustering is found from instantaneous temporal slice by temporal stability. Temporal stability confirms the absence of periodicity (b {\it left}: variation of information $v(t,0,\tau)$ ) and regular reorganisations of the community structure of the trade network (b {\it right}: variation of information $v(t,\tau,\tau+1)$). Temporal stability foxily identifies abrupt changes in the community structure that correspond to major historical events of the 20$^{th}$ century. The date of said events being easily read on the abscissa of (b {\it right}). The variation of information is computed from the optimal partitions.} \label{fig::cow_stability}
\end{figure}

\paragraph{Face-to-face interaction network}
The first real network we analyse is the temporal face-to-face network of children in a primary school described in \cite{Stehle:2011es}. 
The connectivity dynamics of this dataset resembles that of the toy-model, because children's interactions mostly take place within their class and are punctuated by inter-class interactions during breaks and lunch. 
We repeated the analysis above for one day of contacts and were able to uncover the structures described in \cite{Stehle:2011es}.
We find that, for short time-scales $\tau$, the stability curve obtained for $\Delta$-partition with $\Delta=1h$ has a stability score markedly higher than those of partitions obtained over smaller window sizes (Fig. \ref{fig::sociopatterns_stability}a {\it bottom}), which potentially contain more information in this $\tau$ interval.
This reflects the fact that the organization of the children's schedule is organised in hourly periods with short breaks in between.
Hence, the dynamics below that time-scale does not bring extra information.
Like in the toy model, the variation of information $v(t,0,\tau)$ for the optimal partition (Fig. \ref{fig::sociopatterns_stability}b right) shows that the original information is almost completely lost after two hours. 
Since during lunch break children are gathered together and can interact freely across classes, we expect the the lunch break to produce a discontinuity in 
While the temporal stability of the optimal partition is the highest at all delays $\tau$ as expected (Fig. \ref{fig::sociopatterns_stability}a bottom),
$v(t,\tau,\tau+1)$ shows non-trivial patterns which were absent in the toy model example. 
This indicates that, although the global structure of the children's communities changes slowly in time, it can fluctuate from instant to instant considerably more in comparison to the toy model case, highlighting the presence of the social structures beyond that of the classroms.\\
%%%%%%%% 	mucche %%%%%%%%%%
\paragraph{International Trade Network}
\indent The second real network we consider is the financial trade network over the period 1870-2009  \cite{barbieri2009trading,barbieri2008correlates}. The original dataset is weighted and directed. In order to make it consistent with our binary and symmetric null model, we needed to choose a local threshold and forfeit directionality (see Appendix C for details). Despite the loss of information due to these simplifications, temporal stability still captures a rich economical phenomenology.
As it is conceptually very different from the Sociopatterns dataset, we expect to unveil a different type of temporal structure.
Indeed, while no periodicity is present, we clearly identify abrupt changes in the community structure between periods of structural stability (Fig.\ref{fig::cow_stability}b).
These structural changes correspond to landmarks of the 20$^{th}$ century's history that had repercussions on the trade network's community structure: for example two uniform regions are present in Fig. \ref{fig::cow_stability}b (left) between the Great War and the late '20s, and from the late '20s to the beginning of World War II. Temporal stability therefore identifies the beginning of the Great Depression in 1929.   
Note also how the regions of uniformity in Fig. \ref{fig::cow_stability}b appear to become shorter proceeding to more recent times and essentially disappear after 1970. 
It is particularly enlightening to look at Fig. \ref{fig::cow_stability}b (right), which confirms the presence of  spells of relative structural calm, interrupted by sudden changes. 
Zooming in the past forty years, we also see more frequent localized signatures, which appear to be in connection with notable crisis, among which: the oil crisis in the '70s, the speculative bubble of the '80s, the Gulf War, the fall of the Communist Block, the banking crises of the early '90s and finally the 2008 financial crisis.\\
%%%%%%%%%%%%%%%%%%%%%%%%%%%%%%%%%%%%%%%%%%
\section{Discussion}
To summarise, we introduced {\it temporal stability}, a novel measure related to the modular structure of temporal networks. There are two aspects to temporal stability: first, it enables to compare the relevance of different time-varying partitions at different time-scales; second, it can be used to find the optimal partition over a time-scale $\tau$ starting at any time $t$.
The main difference between this method and previous ones lies in the conceptually different treatment of the temporal aspect of the problem.
Time is naturally embedded in temporal stability as it is based on random walks co-evolving with the network.
This is in stark contrast with other methods where time is effectively treated as an additional topological constraint and only probed at one time-scale. The outcomes of different methods are therefore different and complementary (see Appendix D for details) for a complete comparison with the Mucha {\it et al.} method \cite{Mucha:2010bz}).
We illustrated the working of temporal stability on different benchmark datasets, showing that it is capable of discerning characteristic time-scales of the dynamics underlying the datasets as well as highlighting single shocks in the systems. As any target function that includes a comparison between a structure and a notion of randomness, the choice of the null model is crucial. In this Article, our choice fell on the activity driven model for two reasons: it constitutes the simplest non trivial null model and it yields an analytical expression for stationary distribution of random walkers on an evolving network.
In turn, this allows to have a closed form expression for the temporal stability.
The equation for temporal stability (\ref{temporal_stability} and \ref{temporal_autocorr}) can be modified with different null models accounting for different constraints, like specific temporal behaviours (e.g. temporal correlations), for example using stationary distributions obtained from simulated ensembles of paths respecting temporal correlations \cite{Barrat:2013wq}, realistic inter-event times distributions \cite{Rocha:2013ji,Rocha:2011bq,Hoffmann:2012fa}, or temporal null models allowing for weighted links. The investigation of the different possible null models will be the object of future work and is subject to data availability.\\
\\
\paragraph*{Acknowledgements.} The authors acknowledge C. Cattuto and A. Barrat for fruitful discussions and for privileged access to the Sociopatterns data used in this paper. GP is supported by the TOPDRIM project funded by the FET program of the European Commission under Contract IST-318121.I.D. PE is supported by a PET Methodology Program Grant from the MRC UK (Ref G1100809/1). 
%%%%%%
\onecolumngrid
\appendix
\section{Activity driven model description} 
The activity driven network model is a data-driven generator for random temporal networks \cite{Perra:2012fi}. 
In particular, it uses a node's measured activity potential to represent its likeliness to create new links and thus forming its dynamics. 
Given a network of $N$ nodes, assign to each node $i$ an activity rate $a_i$, which is the probability per unit time that node $i$ will create a new link to another node. 
Usually,  the activity is given in the form $a_i = \eta x_i$, where $x_i$ is the activity potential. 
This is done because it allows to control the average number of active nodes in the network through $\eta$. 
The generation of the network snapshots proceeds as follows:
\begin{enumerate}
\item At time $t$, start with a network $G_t$ with $N$ disconnected nodes;
\item With probability $a_i$ node $i$ creates $m$ links to nodes randomly selected with uniform probability. 
\item At the following step, remove all edges and iterate. 
\end{enumerate}
The simplicity of the model allows to calculate analytically a number of properties of the resulting network. 
For example, given a distribution $F(x)$ for the activity potential, the average degree at a given time is given by 
$\label K \rangle_t = \frac{2E_t}{N} = 2m \eta \langle x \rangle$ and the degree distribution of the integrated network after $T$ steps for small time and network size is
$P_T(k) \sim F[\frac{k}{Tm\eta}]$. 
More interestingly for our purposes, the activity driven model allows to describe the asymptotic distribution of a random walker coevolving with the network \cite{Perra:2012ek}. 
It is in fact possible to write the probability of a random walker to be in node $i$ at time $t$ as:
\begin{equation}\label{master}
P_i(t+\Delta t) = P_i(t)\left [ 1- \sum_{j\neq i} \Pi^{\Delta t}_{i\to j} \right] + \sum_{j\neq i} P_j(t) \Pi^{\Delta t}_{j\to i }
\end{equation}
where $\Pi^{\Delta t}_{i\to j} $ is the propagator from $i$ to $j$ over time $\Delta t$. 
In the $\Delta t\to 0$ limit the propagator can be written as $\Pi^{\Delta t}_{i\to j} \sim \frac{\Delta t}{N} (a_i +m a_j)$ and by grouping nodes in activity classes, we can write the equation for the probability $W_a(t)$ of finding the walker in a node of activity $a$ at time $t$ as:
\begin{eqnarray}
\frac{\partial W_a (t)}{\partial t} = -a W_a(t) + amw -m\langle a \rangle W_a(t) \nonumber \\
 + \int a' W_{a'} F(a') da' 
\end{eqnarray}
where $w = W/N$ is the density of walkers in the network. 
Looking for the stationary state of the previous equation, one obtains equation \ref{activity_walker_distribution}. 

\section{Method workflow}
In this Appendix, we describe the workflow to follow to obtain the temporal partition with the optimal temporal stability:
\begin{enumerate}[1.]
\item Extract the adjacency time-series $\{\mathbf{A}_0,\mathbf{A}_1, \ldots, ,\mathbf{A}_T\}$ of the system.
\item Calculate the activity $a$ for each node.
\item Calculate $\omega_{a}$ for each class of activity.
\item Go through each possible pair $[t,\tau]$ and compute the transition matrix $\mathcal{M}_G(t,\tau)$.
\item Compute the modularity matrix $\mathbf{B}(t,\tau) = \left [\mathbf{\Omega}\mathcal{M}_G(t,\tau) - \omega^T \omega \right]$.
\item Find the partition $\mathbf{H}^T(t)$ that optimises $\mathbf{H}^T(t)\mathbf{B}(t,\tau)\mathbf{H}(t)$ using any modularity optimisation algorithm (for example the Louvain method \cite{Blondel:2008do}).
\item Finally average over $t$ to find the temporal stability $r_{\tau, \mathscr{H}\{t\}}  =  \left \langle  \Tr R_{t',\tau}  \left [G\{t\},  \mathscr{H}\{t\} \right ] \right \rangle_{t'}$.
\end{enumerate}
The code to perform this is available at  \url{https://github.com/lordgrilo/TemporalStability}.

\section{Datasets}
\begin{figure}[h]
\centering
\begin{subfigure}[t]{.8\textwidth}
\includegraphics[width=\textwidth]{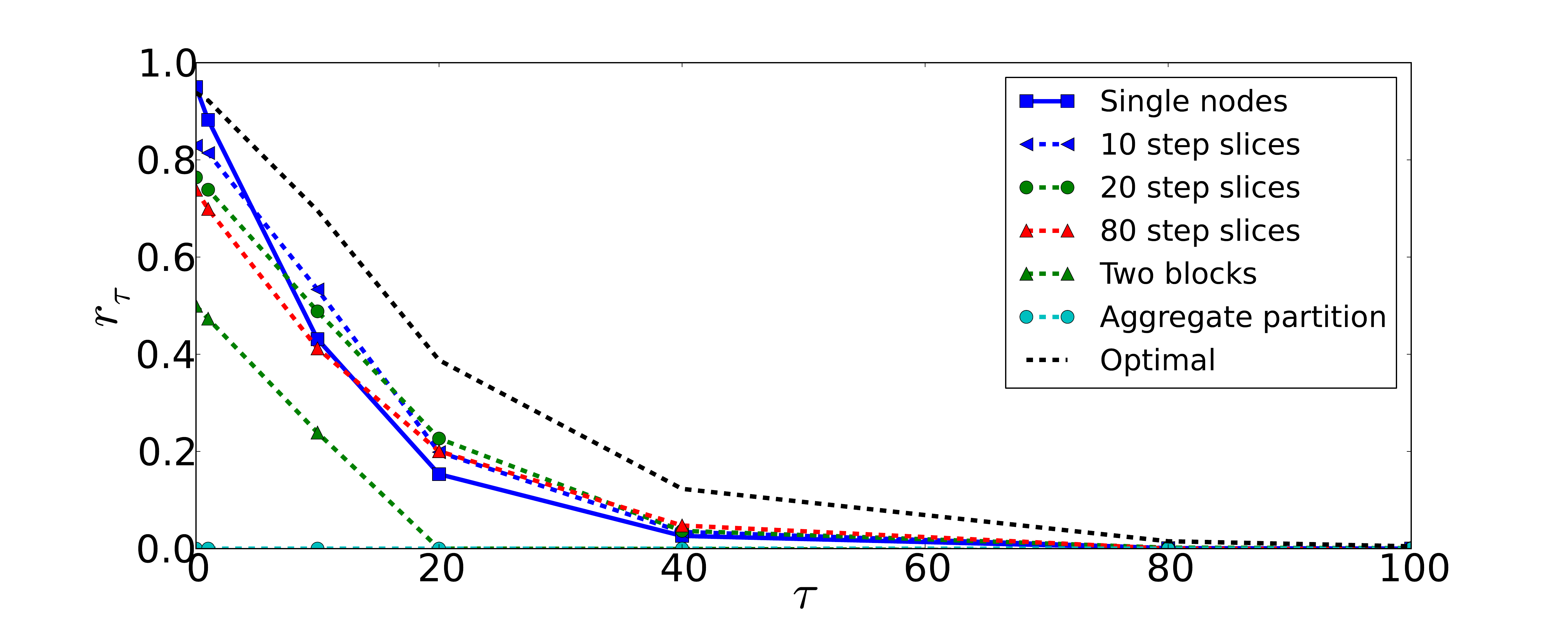}
\caption{Temporal stability $r_\tau$ curves for a selection of $\Delta$-partitions against the best temporal clustering obtained from Eq. (7) in the main text. We show here an extended version of the plot from the main text, with additional $\Delta$-partitions, the partition obtained for the aggregated network and the trivial partition where each node is in its own community at all times. Note that the aggregated partition has zero stability for all $\tau$; this is expected as in this case all nodes are in a single partition by construction.}
\end{subfigure}
\begin{subfigure}[t]{.8\textwidth}
\includegraphics[width=\textwidth]{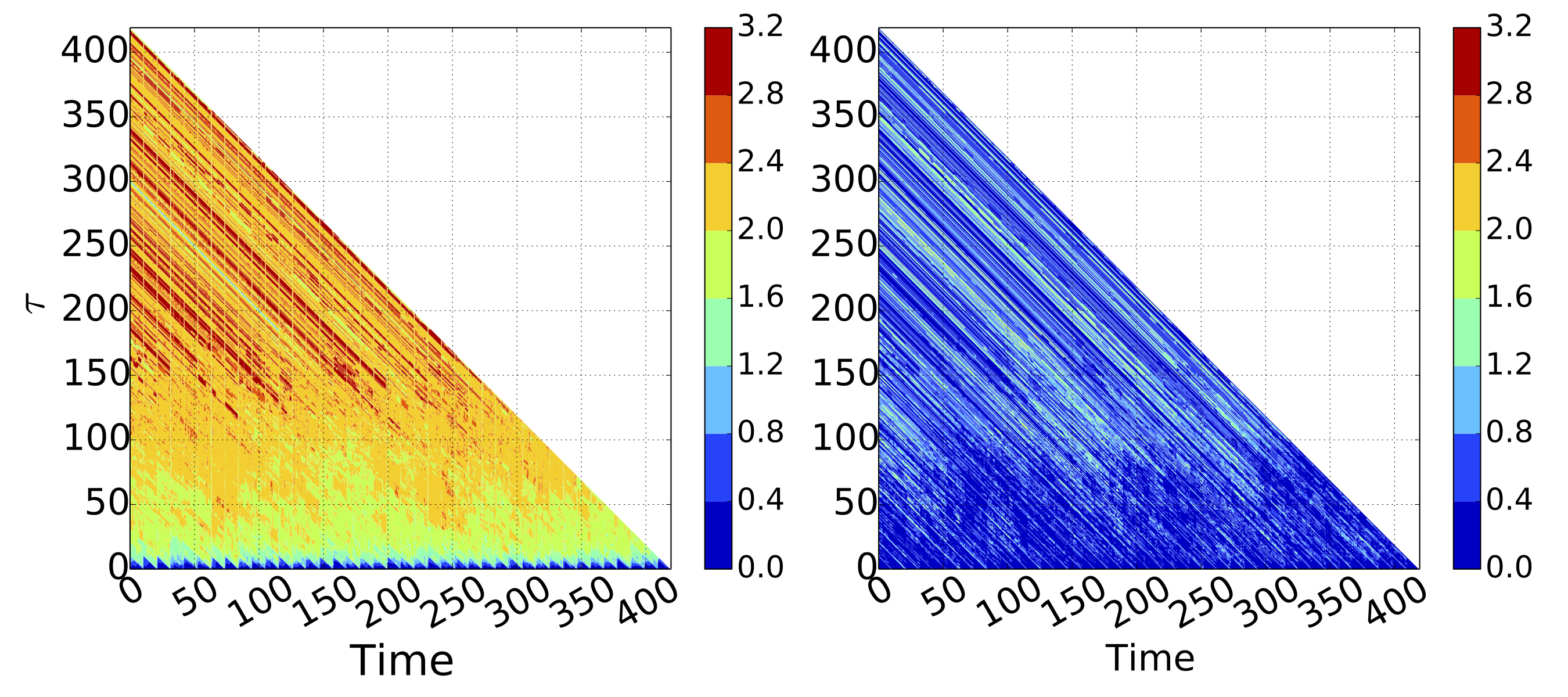}
\caption{The variation of information $v(t,0,\tau)$ ({\it left}) and $v(t,\tau,\tau+1)$ ({\it right}). The periodic nature of the
model is captured very clearly in the saw-tooth patterns for small $\tau$ values, corresponding to the mixing behaviour of the random walker moving on the network. Since there is no special structure in addition to the two blocks, we do not see any interesting features for $v(t,\tau,\tau+1)$ apart from some noise due to fluctuations within the two blocks.}
\end{subfigure}
\caption{{\bf Two block model}.} \label{fig::two_block_stability}
\end{figure}

\textbf{Two blocks model.} The model consists of two sets of nodes of cardinality $N$ and $N'$.
We assign to each node $i$ a value $b_i \in \{0,1\}$ depending on whether they belong to the first or the second block. 
We create a temporal network with $M$ cycles by realizing graph snapshots as follows: 
\begin{enumerate}[i.]
\item  for every snapshot we start with the nodes alone, without any edges;
\item  for a number of steps $T_{in}$ we introduce an edge between nodes $i$ and $j$ with probability $p_{in}\delta(c_i,c_j)$;	
\item  after $T_{in}$ iterations, for $T_{out}$ steps, we introduce  an edge between $i$ and $j$ with probability $p_{out} (1-\delta(c_i,c_j))$. 
\item repeat $M$ times the steps above.
\end{enumerate}
\noindent In particular we are interested in setting the parameters of the block model is such a way as to obtain a time-aggregated network that does not display any obvious community structure. 
This can be done by setting the inter-block linking probability to $p_{out} = \frac{T_{in}p_{in}}{T_{out}}\frac{(N(N-1)+N'(N'-1)}{2NN'}$. For very small values of $M$, fluctuations in the link patterns can prevent the aggregated network from being uniform. However, already for $M\geq 15$ this is not the case anymore. The simulations used in the paper were obtained for $M=20$. \\
In Figure \ref{fig::two_block_stability}, we plot extended versions of the plots of $r_\tau$ and $v(t,0,\tau)$ shown in the main text. Additionally, we show also $v(t,\tau,\tau+1)$ for the two blocks model, which as expected has only trivial signal due to linking fluctuations within the blocks. \\
%%%

\textbf{Face-to-face contact network. } 
The dataset refers to time-resolved face-to-face interactions of 10 teachers and 232 children between 6 and 12 years old. 
Every child was asked to was a small wearable RFID device scanning for other neighboring devices every 20s, corresponding to the resolution of the resulting temporal network.    
Although the original dataset covers two school days from morning to evening, we only analysed on the first day as it is enough to uncover the temporal patterns we are interested in. More details on the exact experimental design can be found in \cite{Stehle:2011es}. 
The aggregated network is available from the Sociopatterns website at \url{http://www.sociopatterns.org/}. \\
As per the two blocks model, in Figure \ref{fig::sociopatterns_stability} we plot extended versions of the plots of $r_\tau$ and $v(t,0,\tau)$ shown in the main text. 
Additionally, we show also $v(t,\tau,\tau+1)$ for the Sociopatterns data. In this case the instantaneous variation of information shows much more structure than the two blocks model in Fig. \ref{fig::two_block_stability}, representing the presence of additional social structure beyond classes. This is especially clear in the time period corresponding to the lunch break, where children are allowed to mingle more freely in the school's cafeteria. \\
\begin{figure}
\centering
\begin{subfigure}{.8\textwidth}
\includegraphics[width=\textwidth]{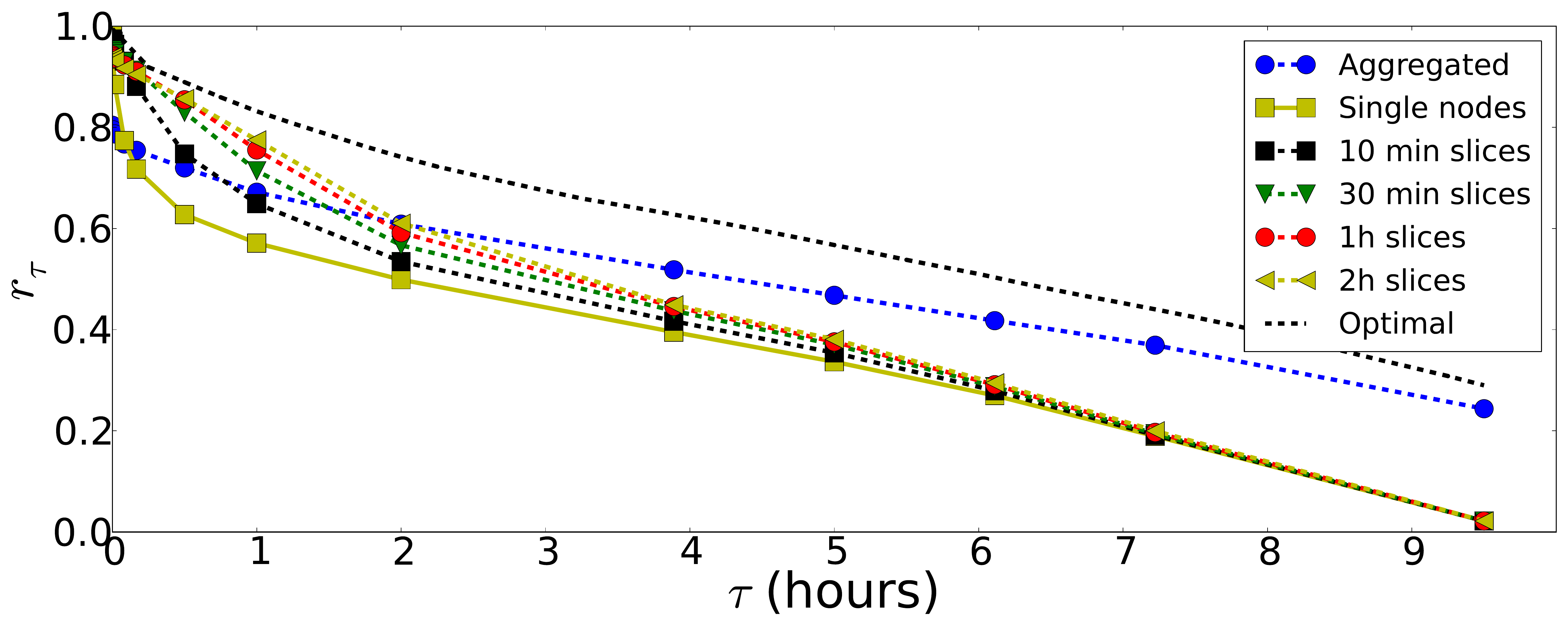}
\caption{Temporal stability $r_\tau$ curves for a selection of $\Delta$-partitions against the best temporal clustering obtained from Eq. (7) in the main text. The partition of the aggregated network develops $r_\tau$ values compatible with the optimal for $\tau >4$ hours: this is due to the fact that over such coarse graining, the effect of the short inter-class breaks and of the lunch break is almost washed away by the relative persistence during lecture hours.}
\end{subfigure}
\begin{subfigure}{0.8\textwidth}
\includegraphics[width=\textwidth]{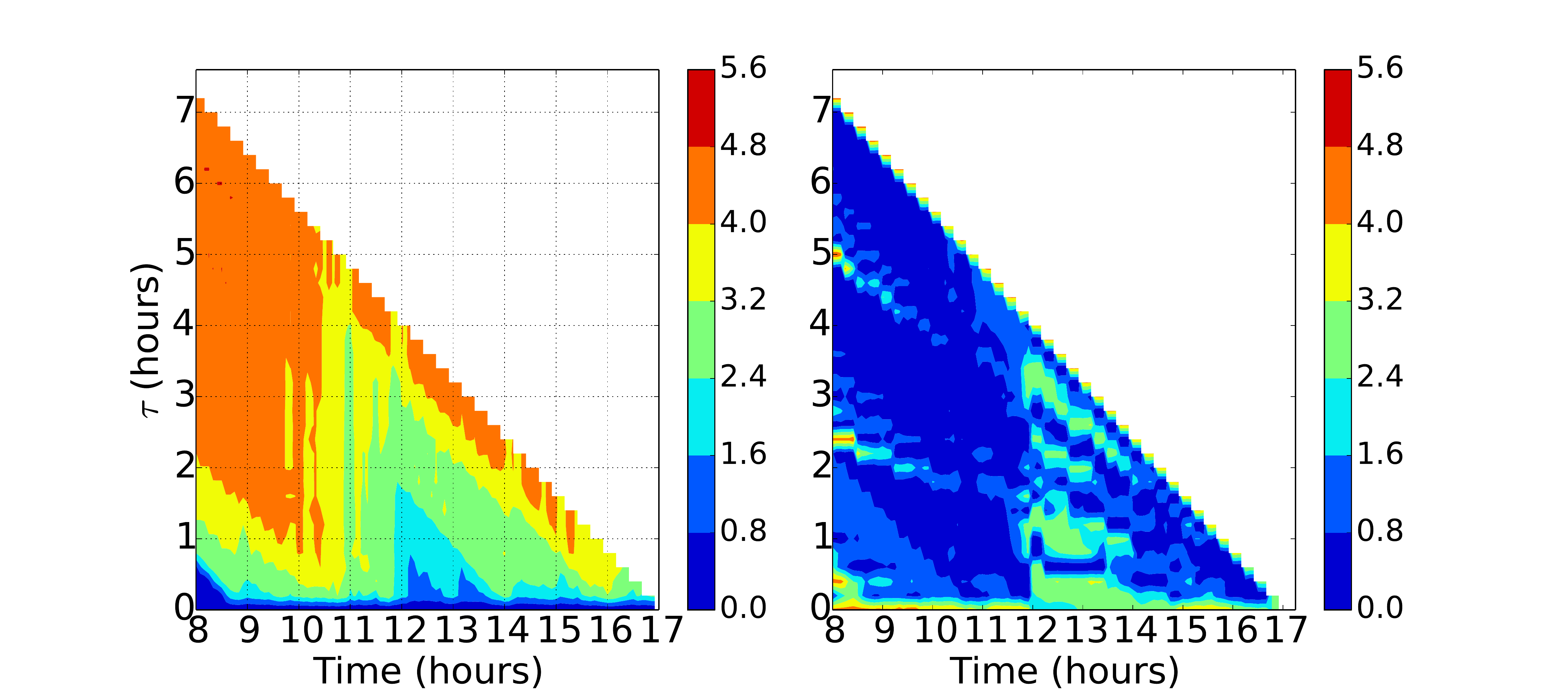}
\caption{The variation of information $v(t,0,\tau)$ ({\it left}) and $v(t,\tau,\tau+1)$ ({\it right}). 
The different dynamics of the temporal network (lecture hours, lunch break etc.) appear naturally and are consistent with the schedule reported in \cite{Stehle:2011es}. In contrast to the lack of structure in Fig.\ref{fig::two_block_stability}b, here we see non trivial information transfer over longer $\tau$ values in $v(t,\tau,\tau+1)$, indicating temporal correlations in the community structure of the children during lunch break.}
\end{subfigure}
\caption{{\bf Sociopatterns data.}} 
\label{fig::sociopatterns_stability}
\end{figure}

\textbf{International Trade Network. }
The dataset consists of trade flows between states for the years 1870-2009 (v3.0) and is publicly available from the Correlates of War website at \url{http://www.correlatesofwar.org/} \cite{barbieri2009trading,barbieri2008correlates}. 
We focused on the dyadic trade network, where each link represents the trade flow between pairs of states in current U.S. dollars. 
The full networks is therefore weighted and directed. 
The activity driven model that we are using as null model however produces binary undirected networks.
Therefore, in order to make the comparison meaningful, we extracted the backbone of the network using the disparity filter  \cite{Serrano:2009ve} with a significance threshold of $\alpha=0.05$, then we symmetrised the network and from there calculated the activity values needed to define the activity driven model. 

\section{Comparison with multislice modularity}
In this section, we illustrate the difference between temporal stability and other temporal partition methods, and focus on multislice modularity \cite{Mucha:2010bz} optimised with the GenLouvain algorithm.

The main difference between these two methods is how time is treated. The approach in \cite{Mucha:2010bz} is a modularity one. They follow the exposition of \cite{lambiotte-2008} construct their generalised modularity on an expansion of the exponential in the Laplacian dynamics for small time. The dynamics they consider on the network is thus effectively a "one-step" dynamics. The time dependence of the system is encoded in the multislice structure as a coupling parameter between the slices which mimics temporal (cor)relation between snapshots. 
Time is therefore introduced as a topological construct.
In contrast, in our approach, we consider the complete dynamics seen by a random walk co-evolving with the network itself, thus embedding the time-evolution of the system in its the description, making it also parameter free.
The two approaches are therefore not only conceptually but also practically different. One has a one time-scale constraint, while the other has access to and uses all time-scales.\\

\begin{figure*}
\centering
\begin{subfigure}[t]{.33\textwidth}
\includegraphics[width=\textwidth]{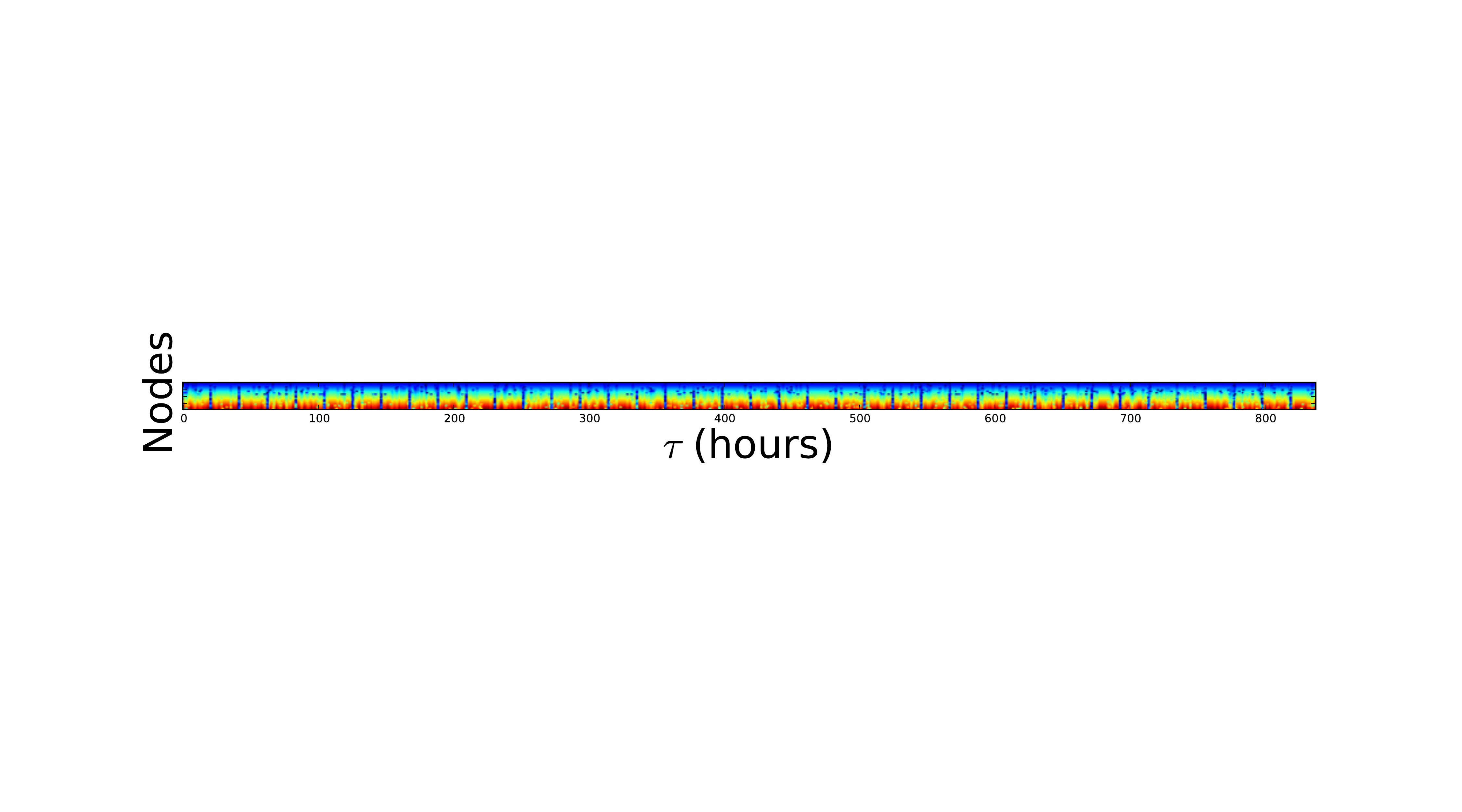}
\includegraphics[width=\textwidth]{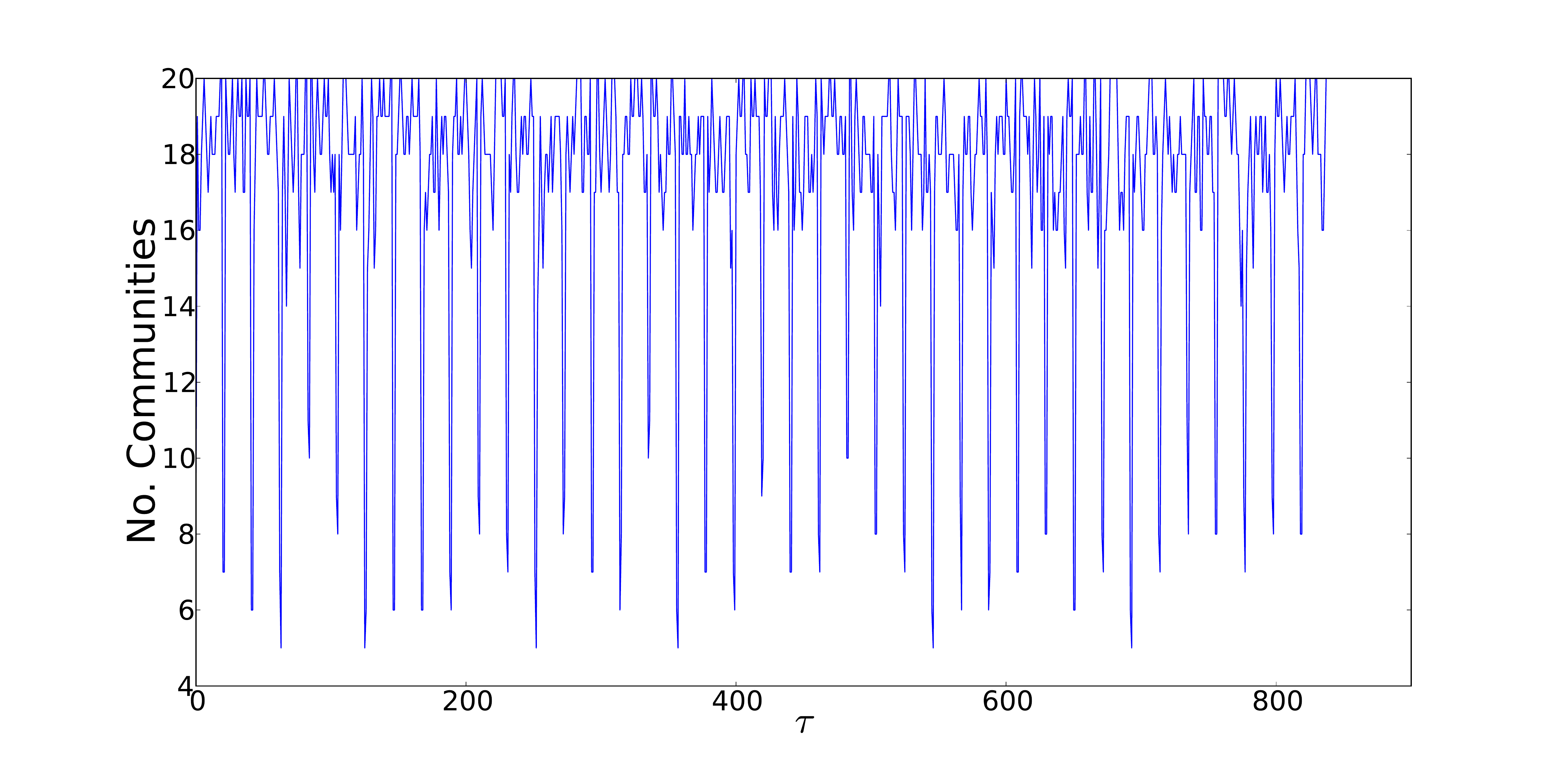}
\caption{Results for optimal partition for temporal stability with $\tau=1$.} \label{fig::two_block_stability_partition_zoomin}
\end{subfigure}
\begin{subfigure}[t]{.33\textwidth}
\includegraphics[width=\textwidth]{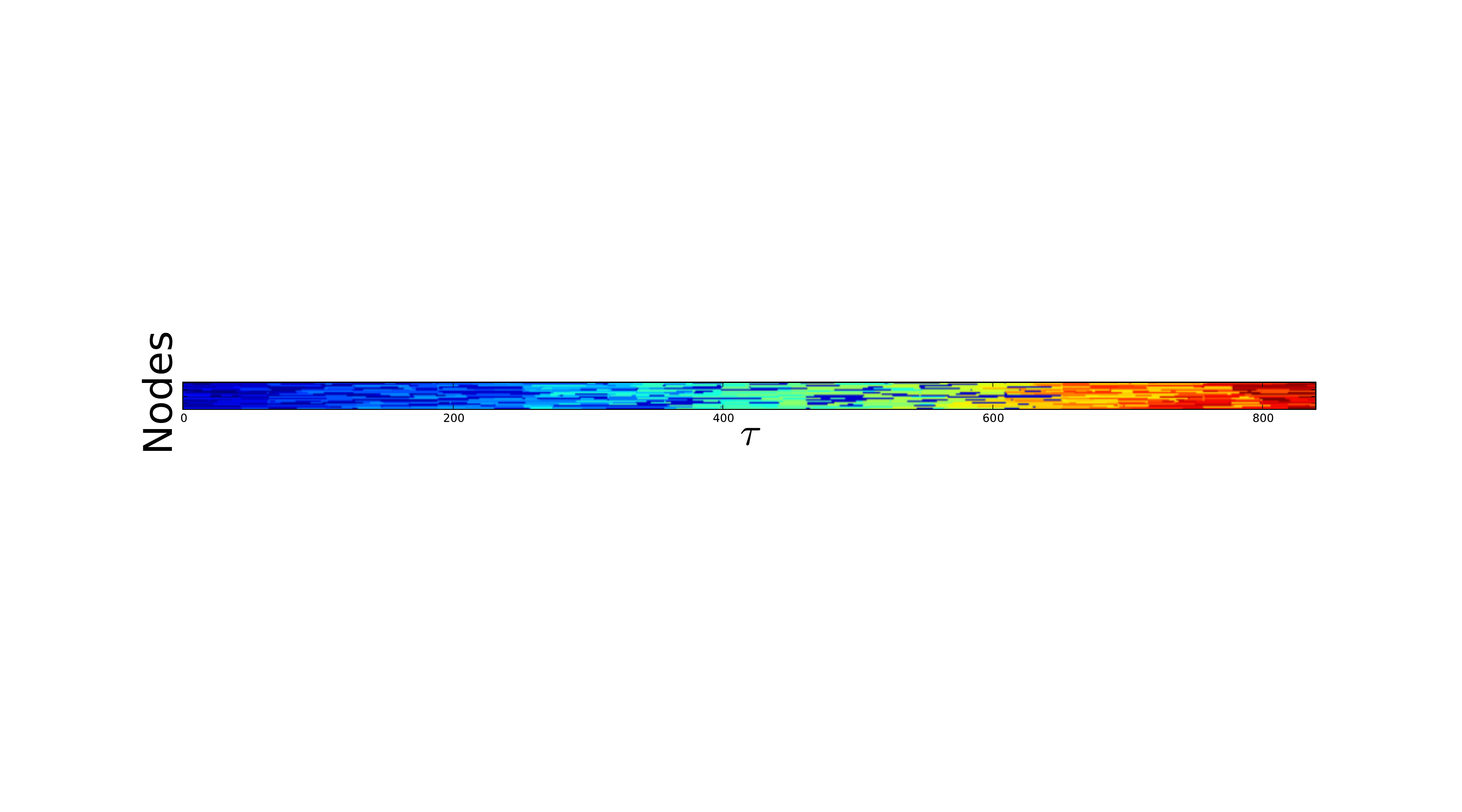}
\includegraphics[width=\textwidth]{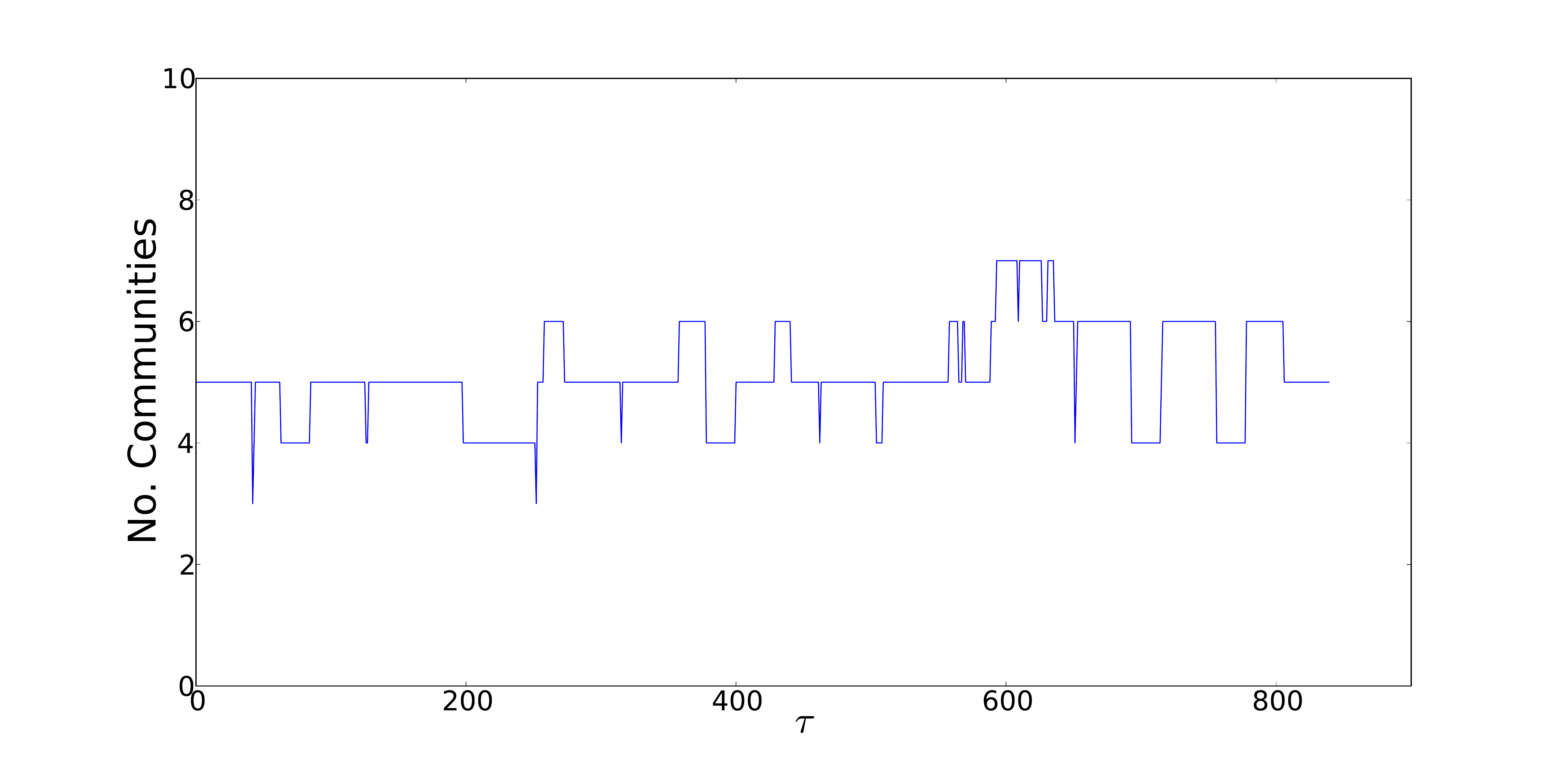}
\caption{Multislice modularity results for $\omega=0.2$.} \label{fig::Two_block_genlouvain_partition_0_2}
\end{subfigure}
\begin{subfigure}[t]{.33\textwidth}
\includegraphics[width=\textwidth]{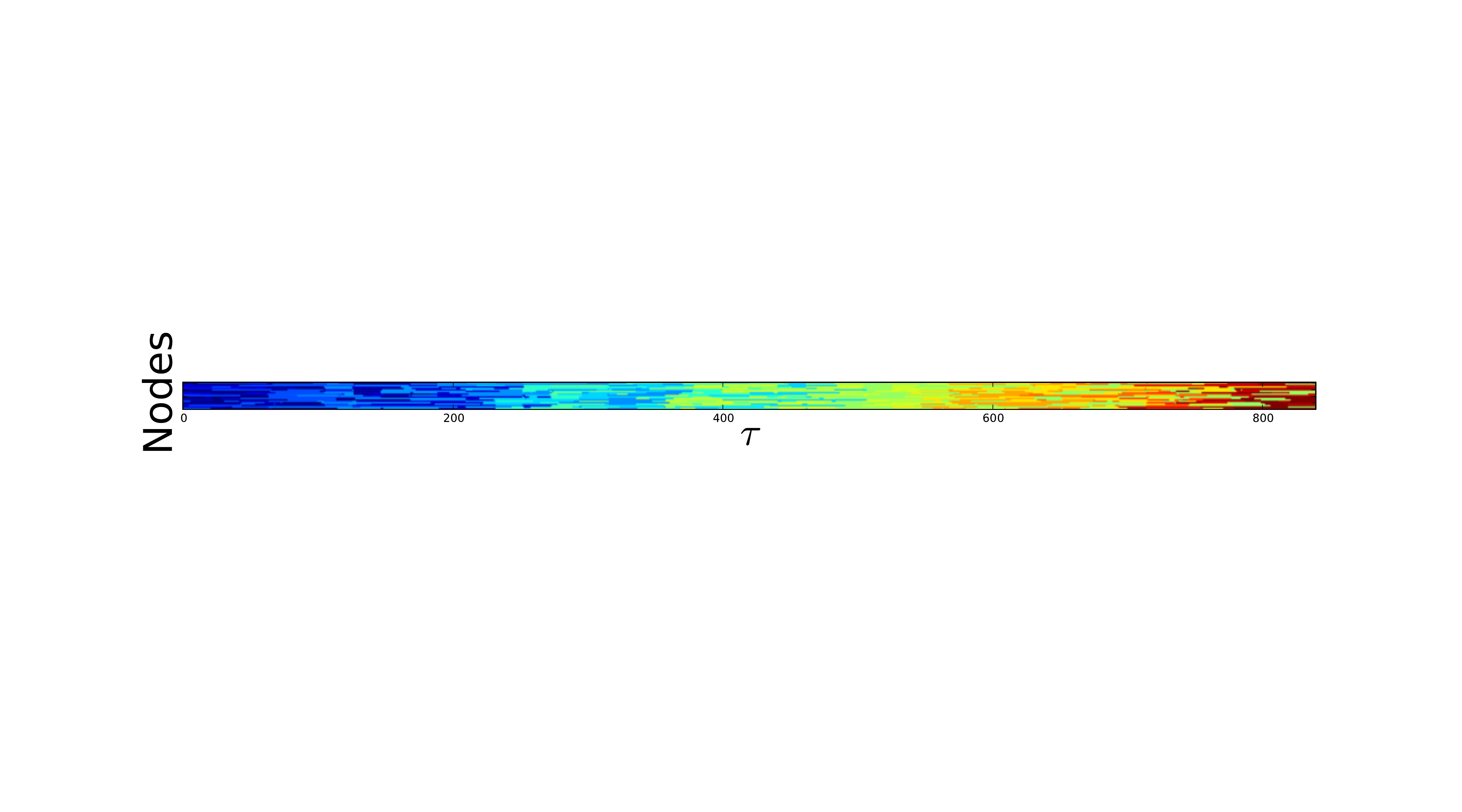}
\includegraphics[width=\textwidth]{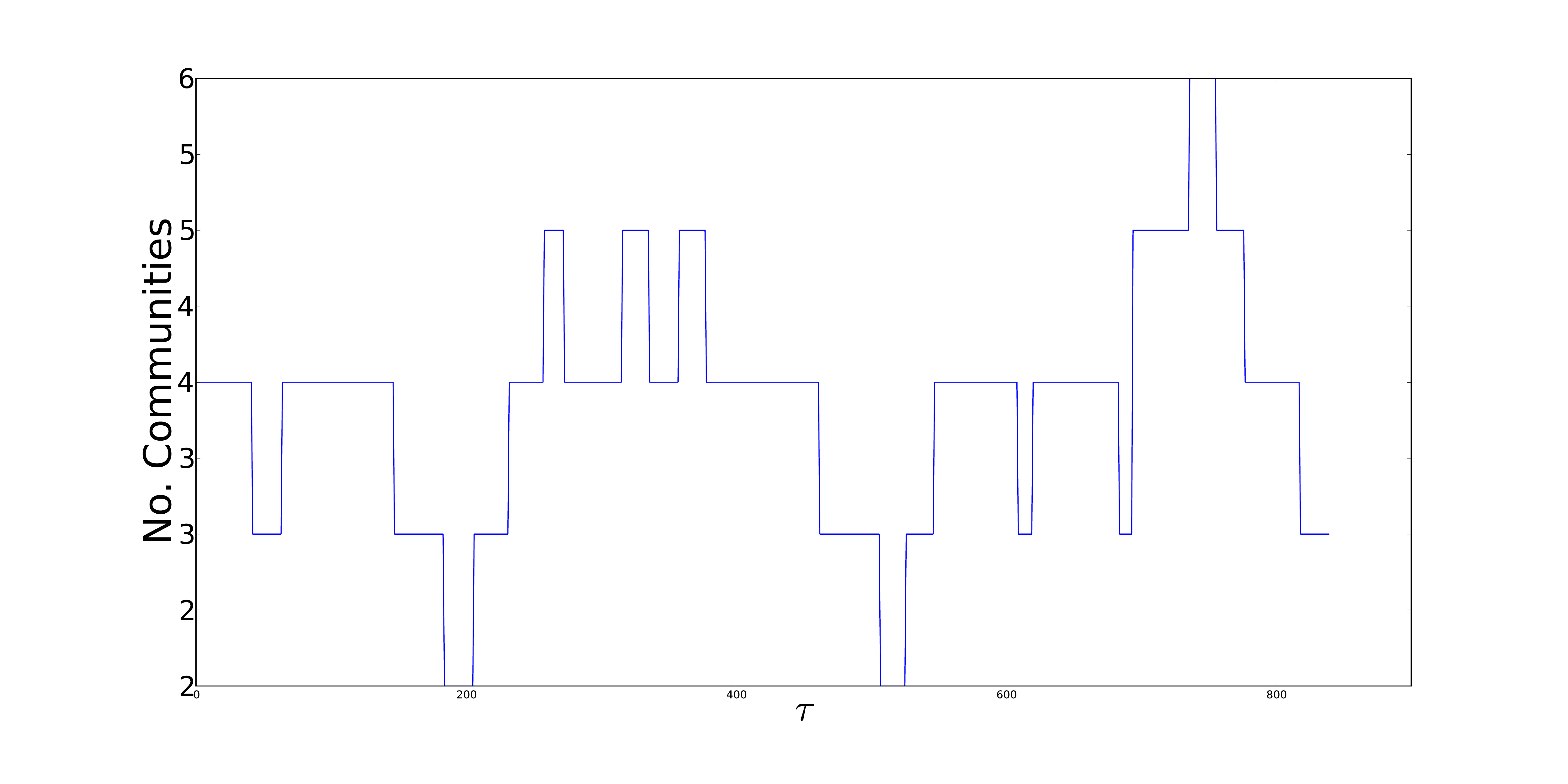}
\caption{Multislice modularity results for $\omega=0.5$.} \label{fig::Two_block_genlouvain_partition_0_5}
\end{subfigure}
\begin{subfigure}[t]{.33\textwidth}
\includegraphics[width=\textwidth]{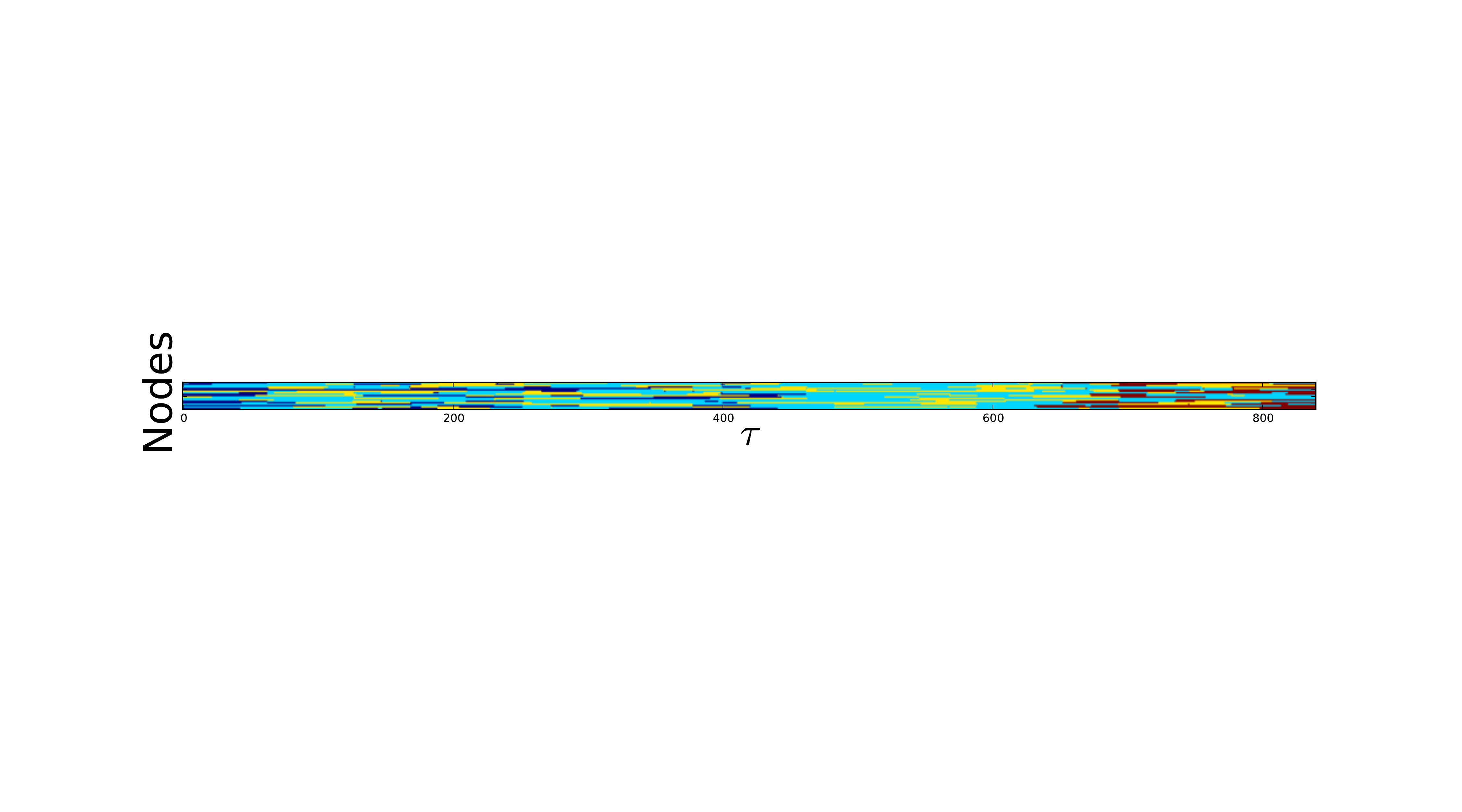}
\includegraphics[width=\textwidth]{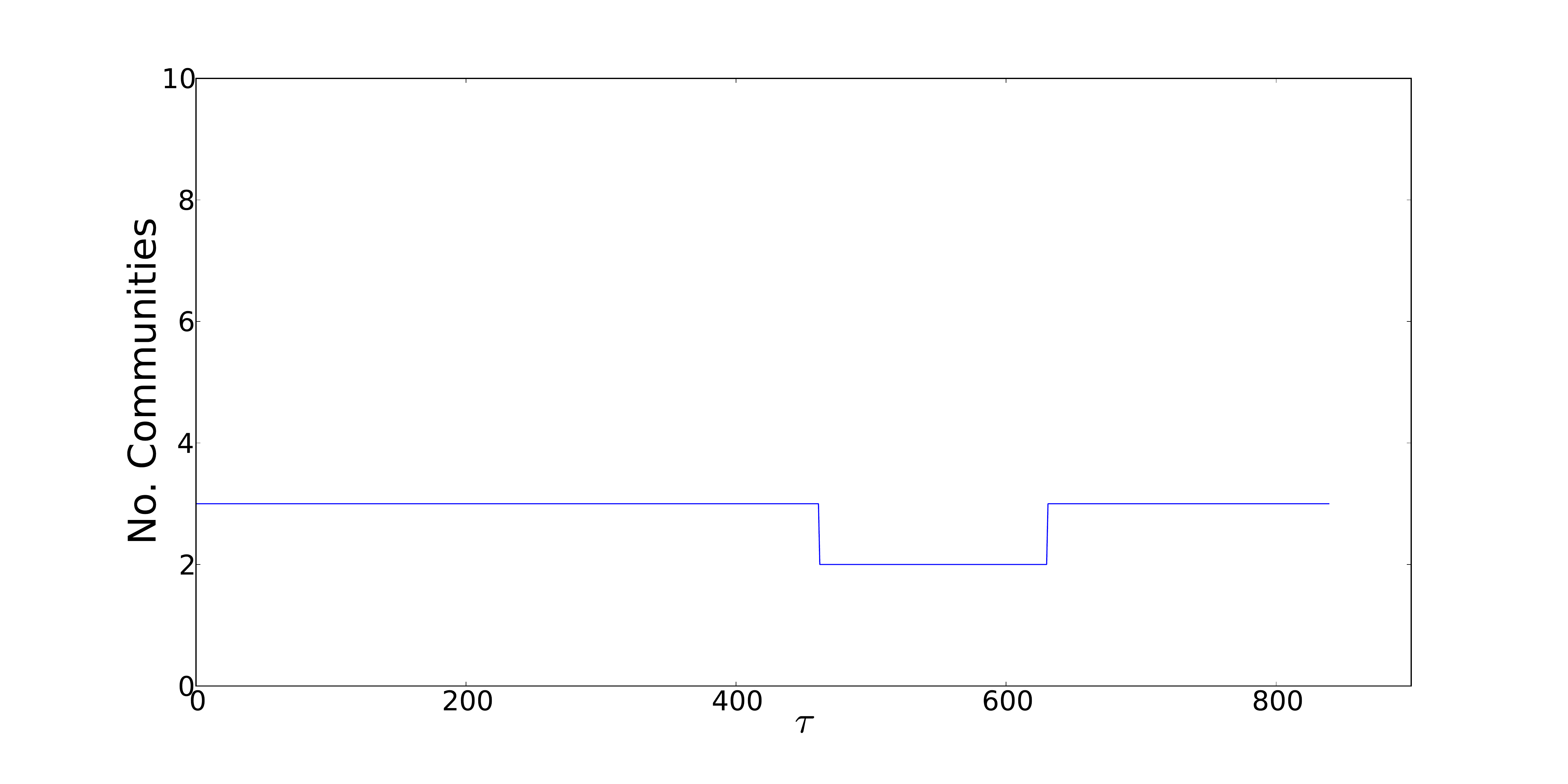}
\caption{Multislice modularity results for $\omega=0.8$.} \label{fig::genlouvain_partition_0_8}
\end{subfigure}
\caption{{\bf Comparison of multislice modularity and optimal stability ($\tau=1$) for the two-blocks model.} Top plots represent community assignment through time, bottom plots show the number of communities at a given point in time.}\label{fig::Two_blockcomparison}

\end{figure*}

\begin{figure*}
\centering
\includegraphics[width=0.5\textwidth]{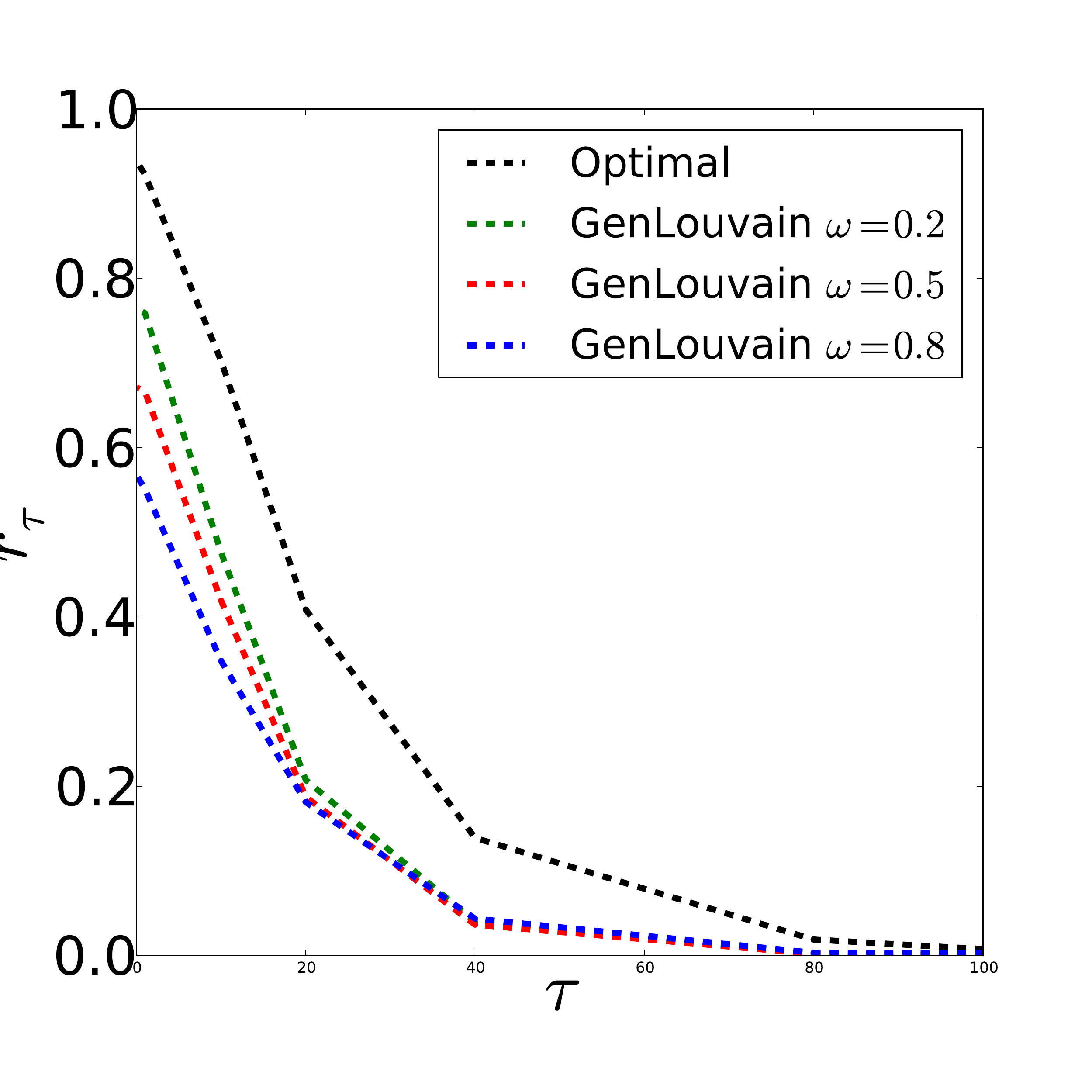}
\caption{\textbf{Comparison of stability values for multislice modularity and optimal partition for temporal stability for the two-blocks model.}} \label{fig::Two_block_comparison_genlouvain_stability}
\end{figure*}

\begin{figure*}
\centering
\begin{subfigure}[t]{.33\textwidth}
\includegraphics[width=\textwidth]{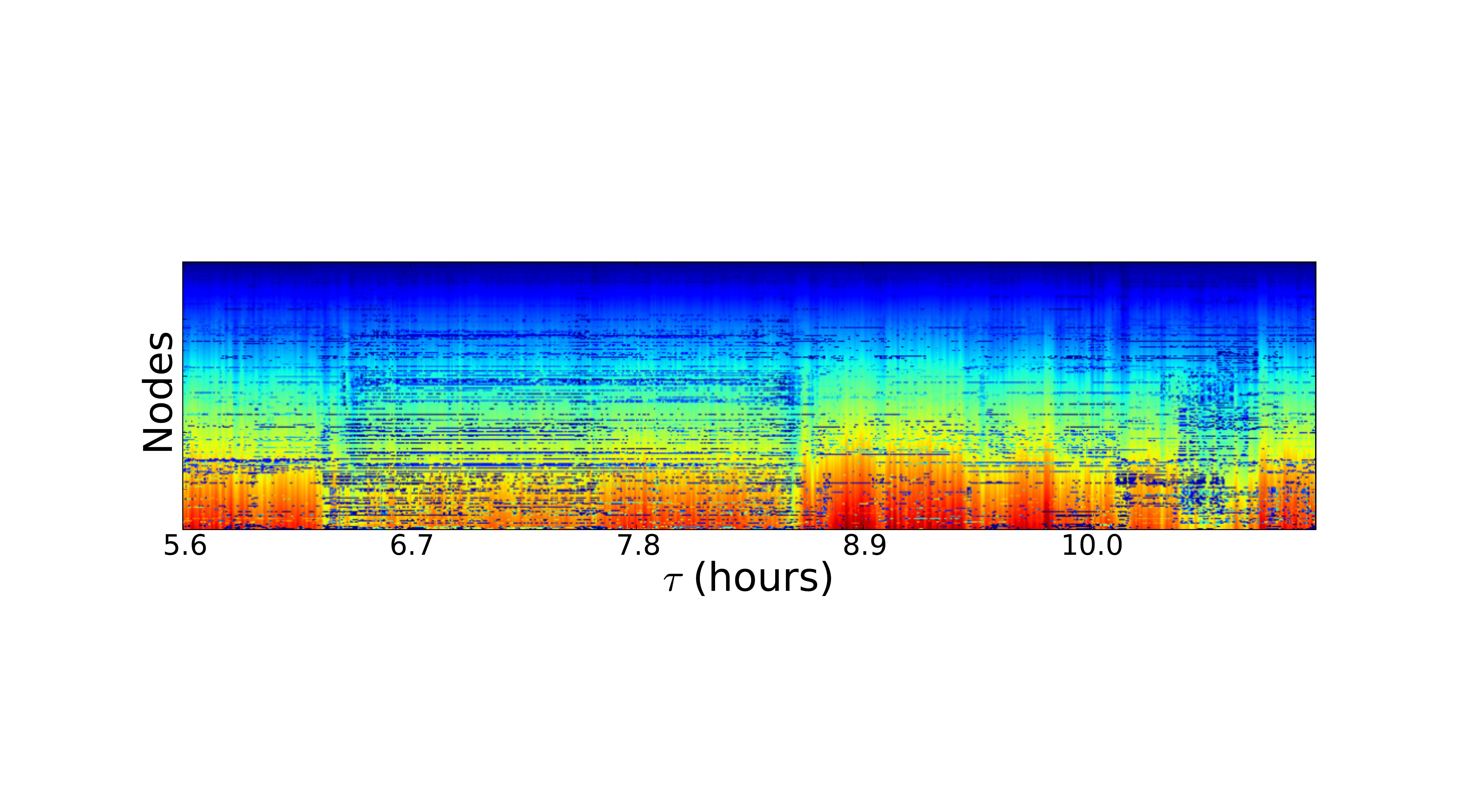}
\includegraphics[width=\textwidth]{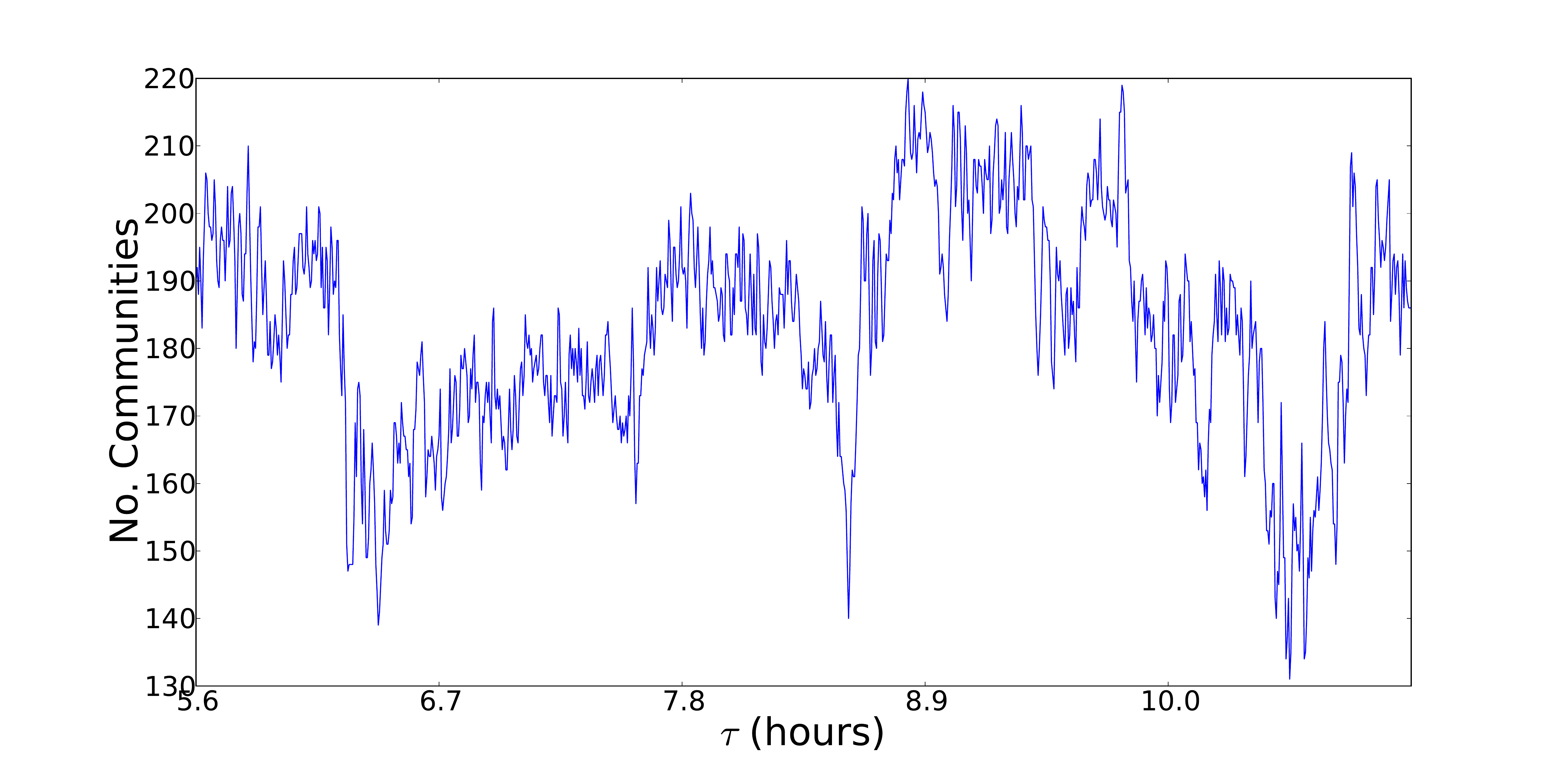}
\caption{Results for optimal partition for temporal stability with $\tau=1$ for the school data.} \label{fig::sociopattern_stability_partition_zoomin}
\end{subfigure}
\begin{subfigure}[t]{.33\textwidth}
\includegraphics[width=\textwidth]{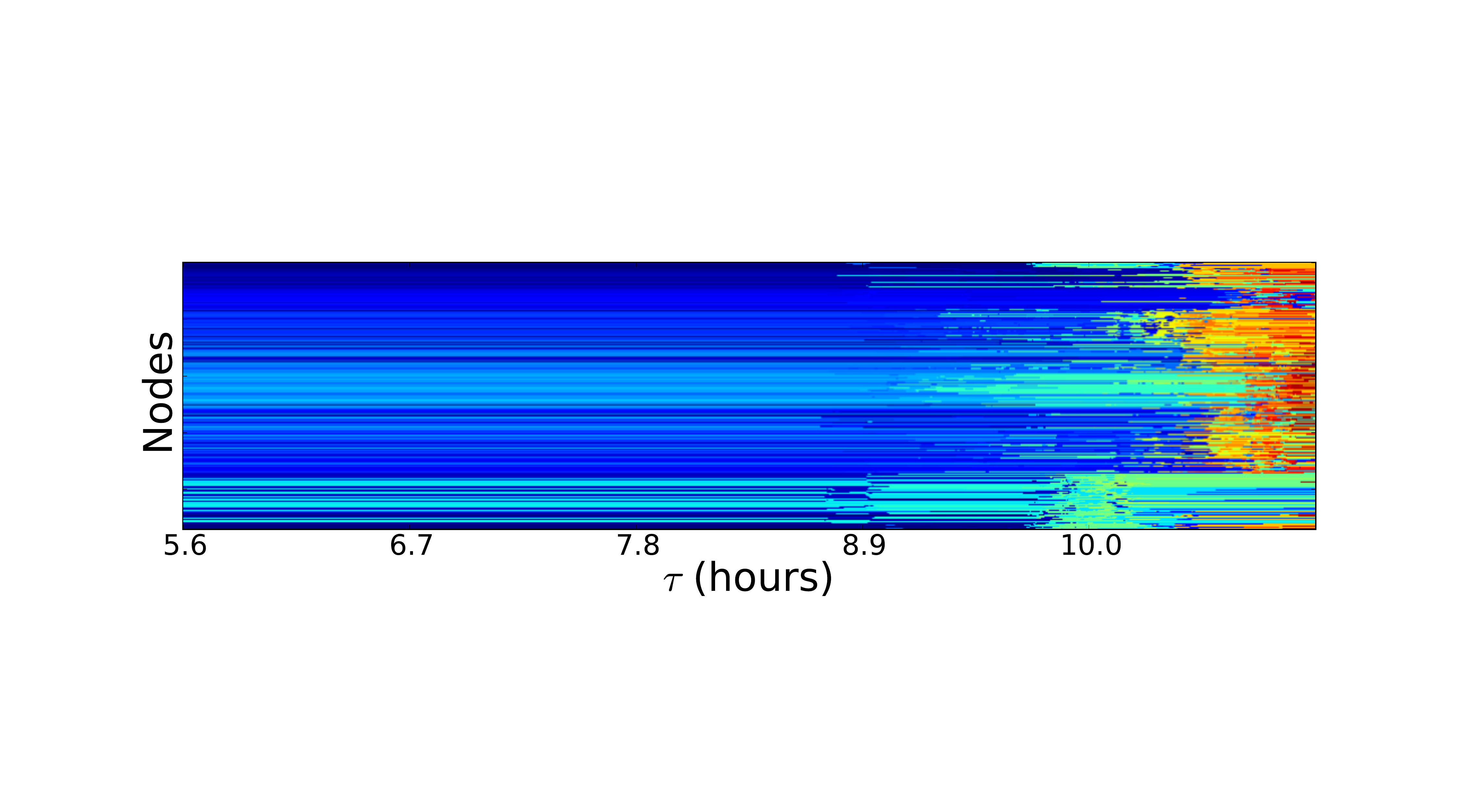}
\includegraphics[width=\textwidth]{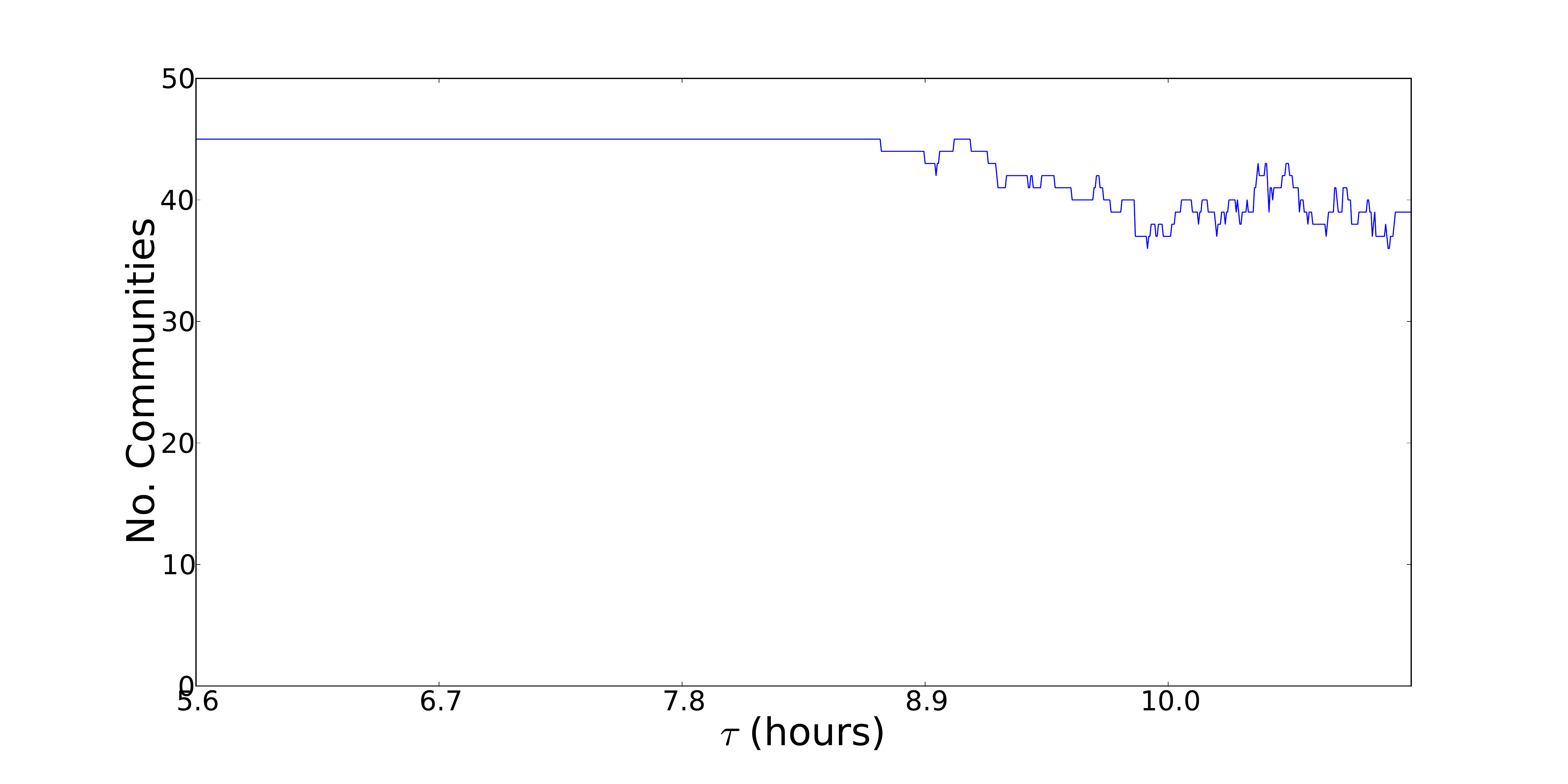}
\caption{Multislice modularity results for $\omega=0.2$.} \label{fig::sociopattern_genlouvain_partition_0_2}
\end{subfigure}
\begin{subfigure}[t]{.33\textwidth}
\includegraphics[width=\textwidth]{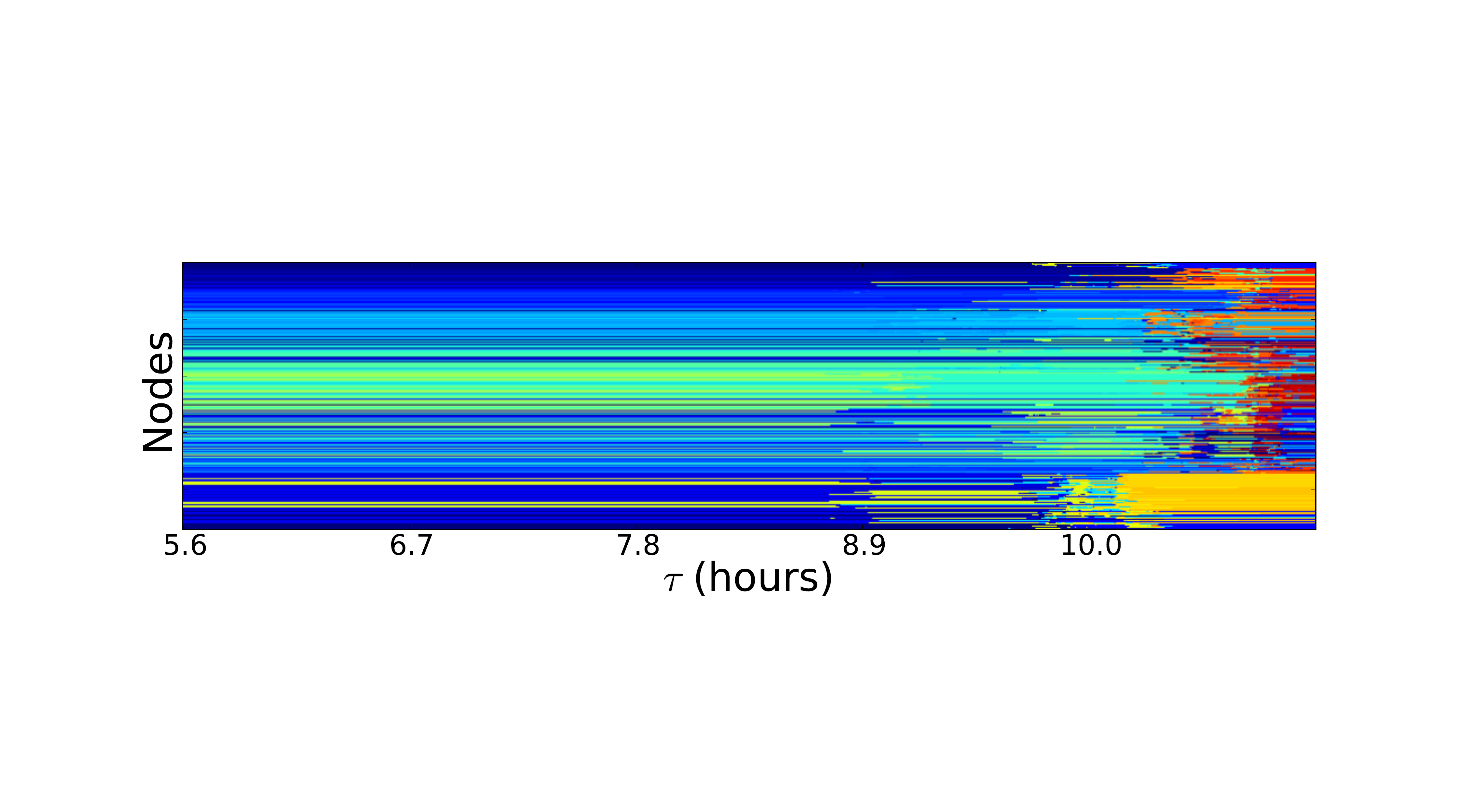}
\includegraphics[width=\textwidth]{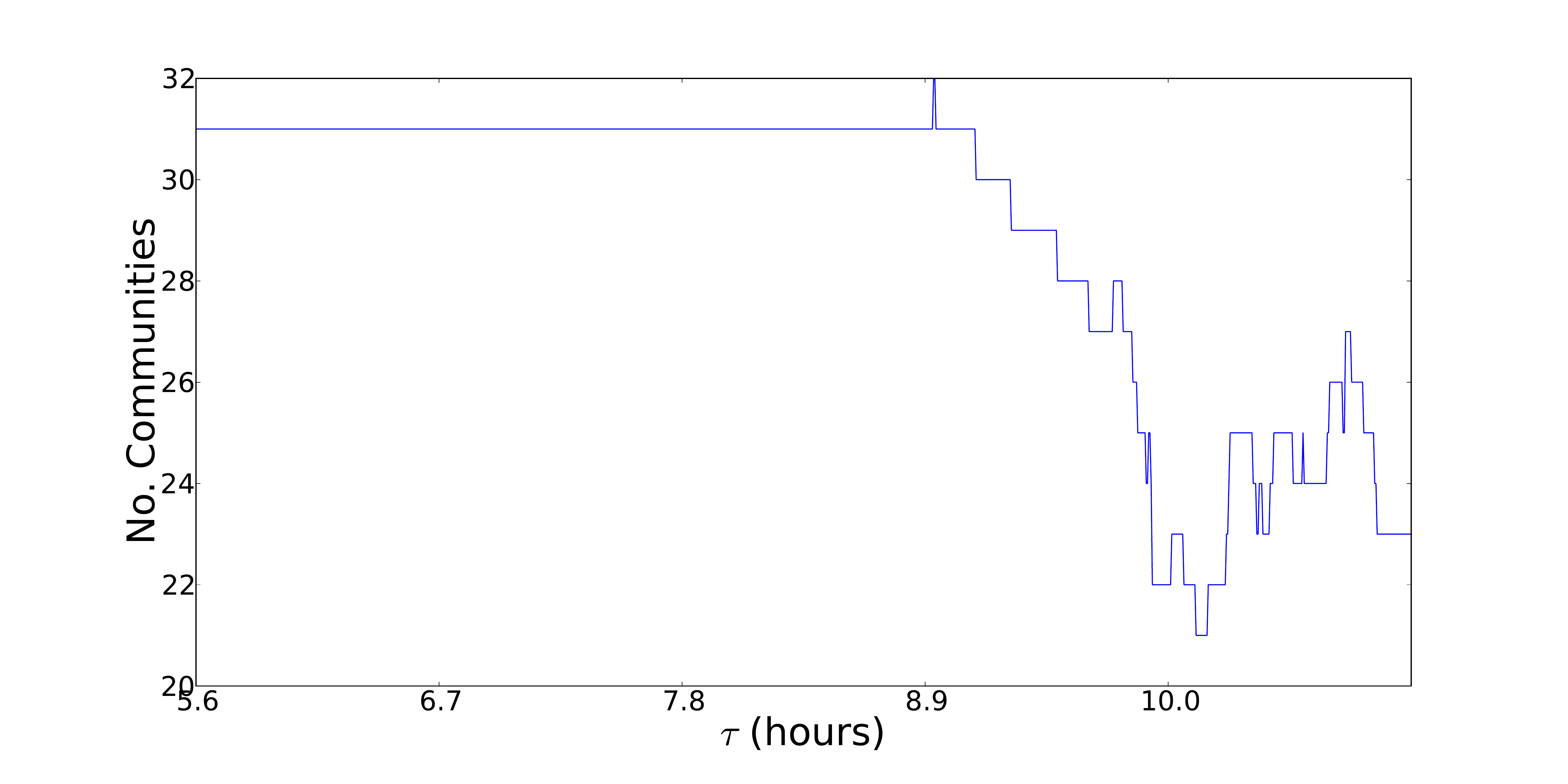}
\caption{Multislice modularity results for $\omega=0.5$.} \label{fig::sociopattern_genlouvain_partition_0_5}
\end{subfigure}
\begin{subfigure}[t]{.33\textwidth}
\includegraphics[width=\textwidth]{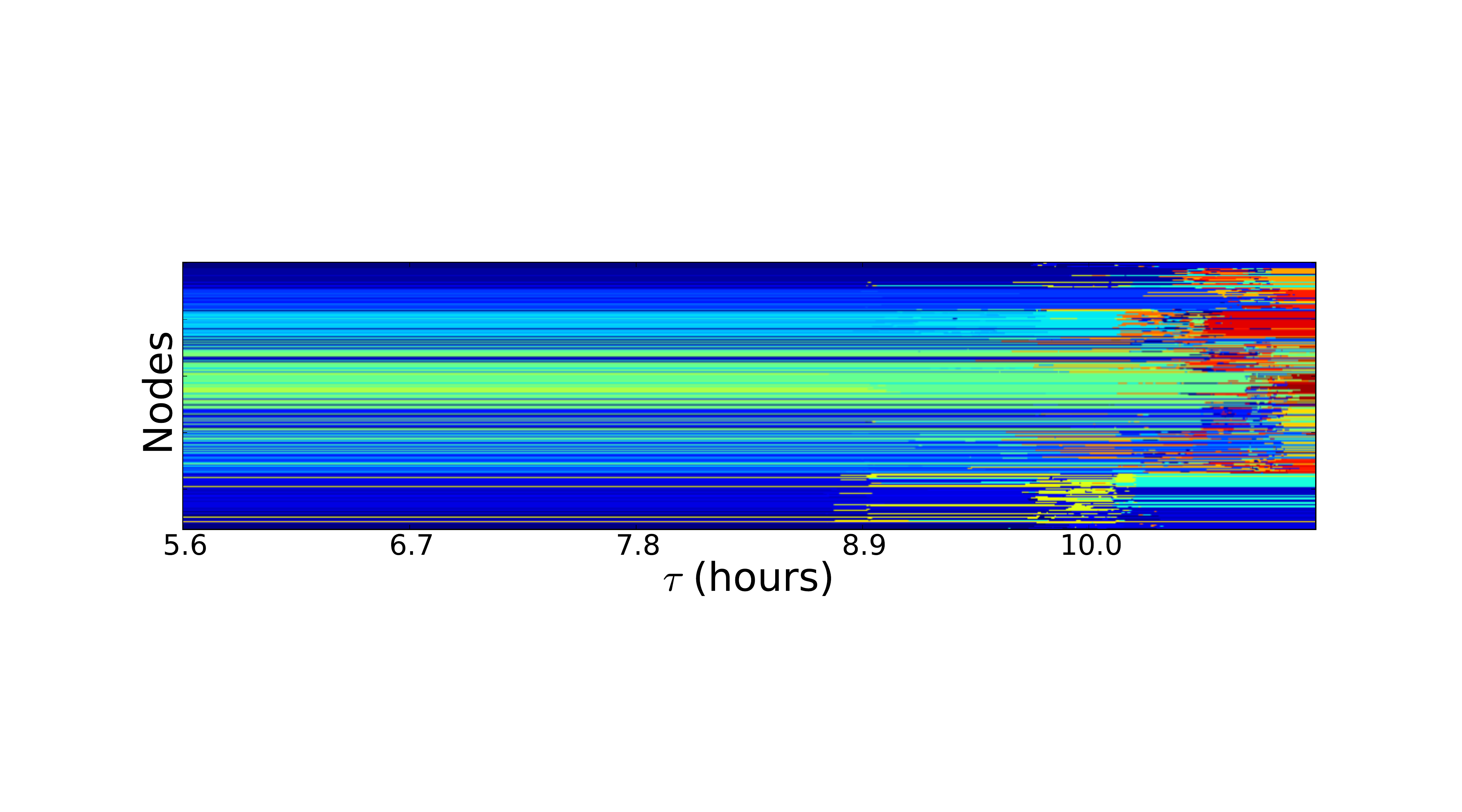}
\includegraphics[width=\textwidth]{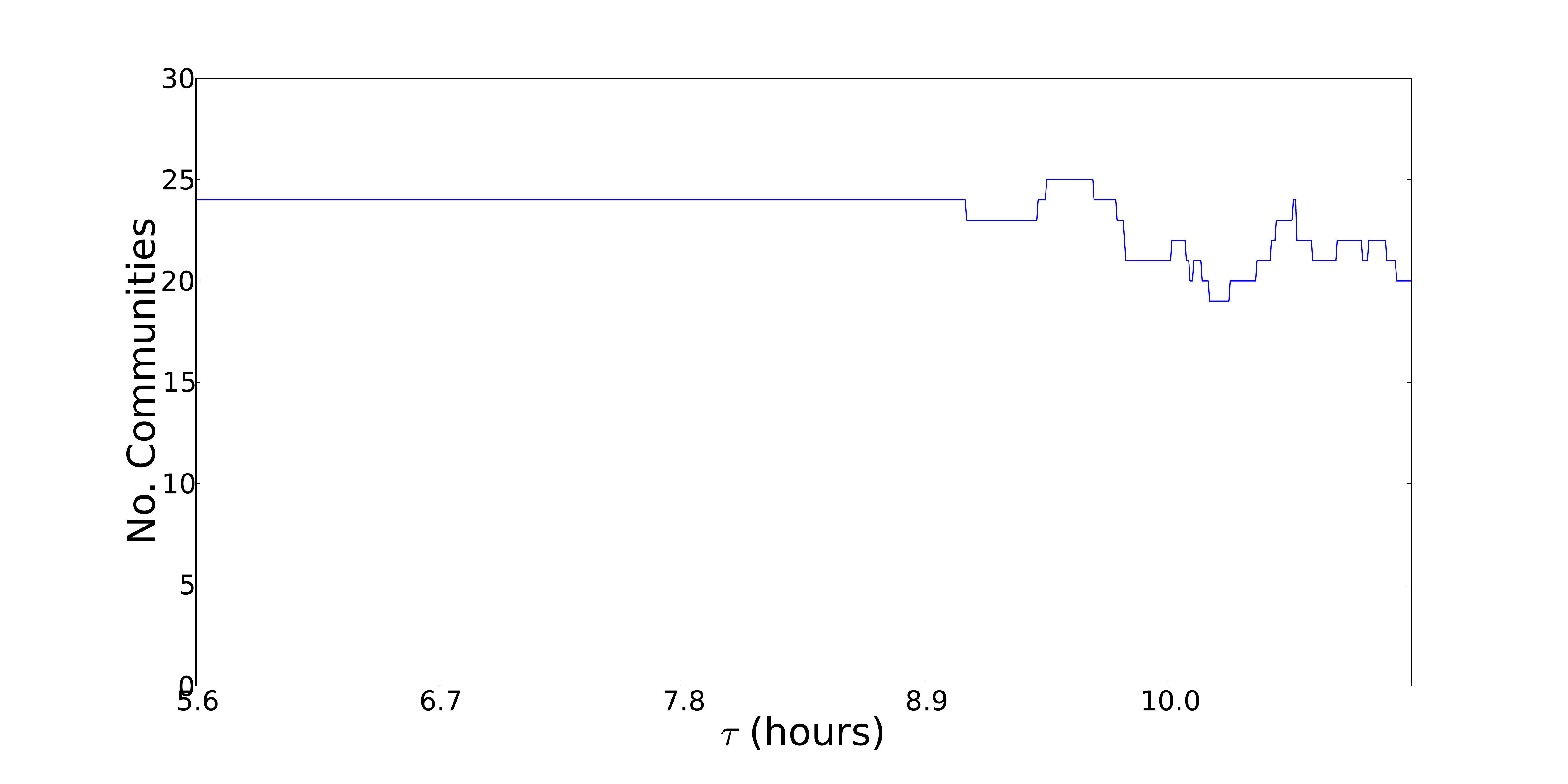}
\caption{Multislice modularity results for $\omega=0.8$.} \label{fig::sociopattern_genlouvain_partition_0_8}
\end{subfigure}
\caption{{\bf Comparison of multislice modularity and optimal stability ($\tau=1$) for the Sociopatterns data.} Top plots represent community assignment through time, bottom plots show the number of communities at a given point in time. The plots shown refer to the central third of the dataset, corresponding roughly to the period 8am-12am. }\label{fig::sociopattern_comparison}

\end{figure*}

\begin{figure*}
\centering
\includegraphics[width=0.5\textwidth]{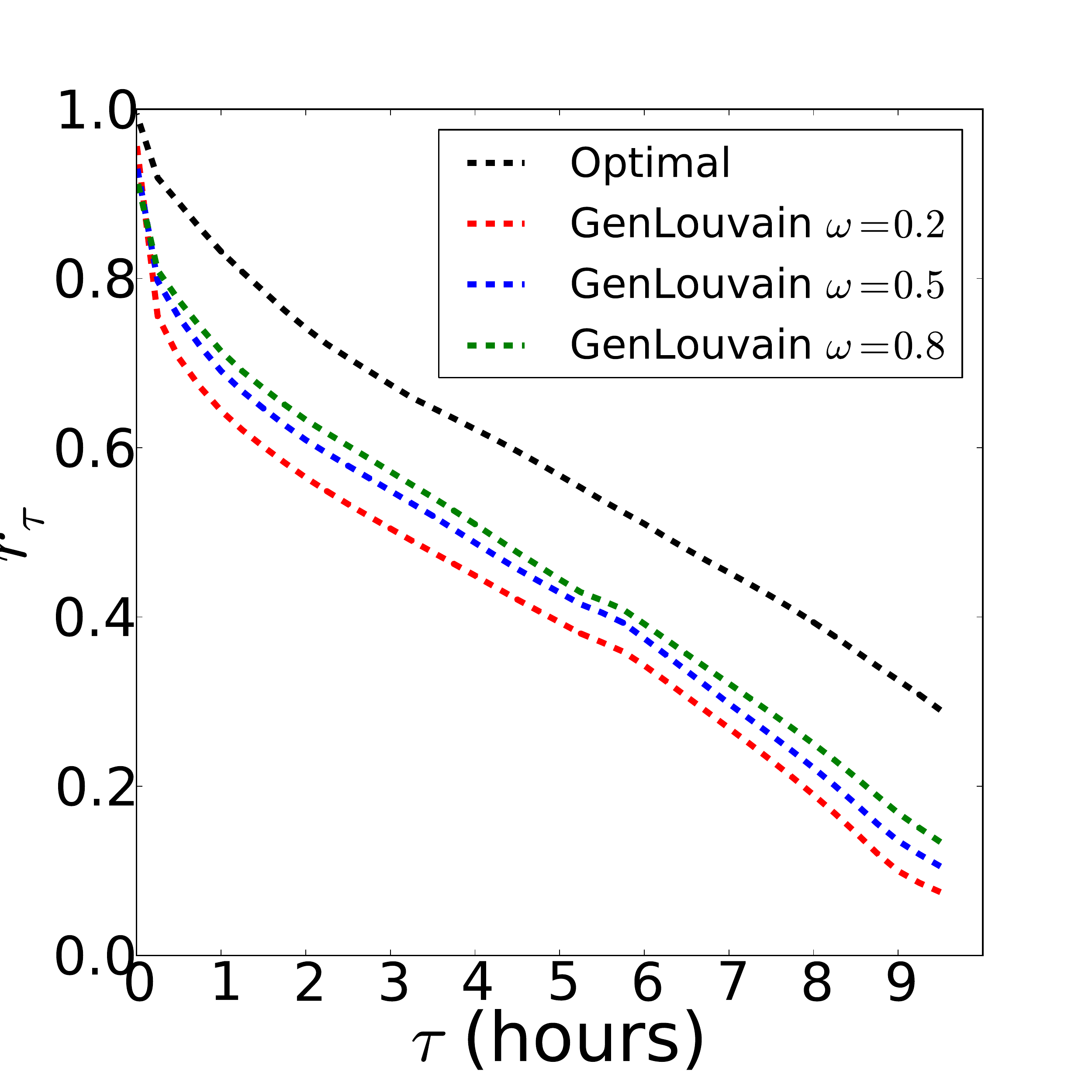}
\caption{\textbf{Comparison of stability values for multislice modularity and optimal partition for temporal stability for the Sociopatterns data.}} \label{fig::sociopattern_comparison_genlouvain_stability}
\end{figure*}

\begin{figure*}
\centering
\begin{subfigure}[t]{0.48\textwidth}
\includegraphics[width=\textwidth]{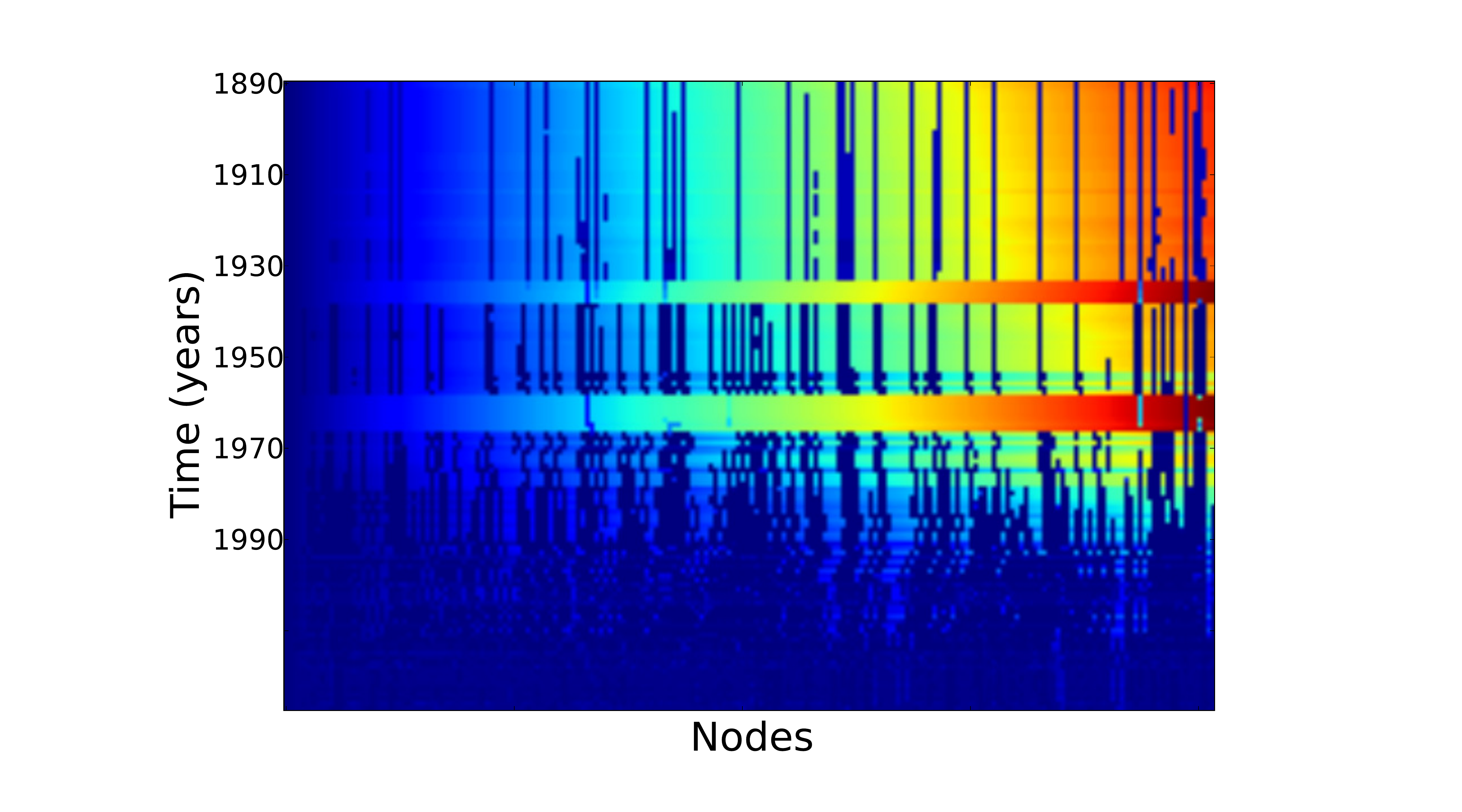}
\includegraphics[width=\textwidth]{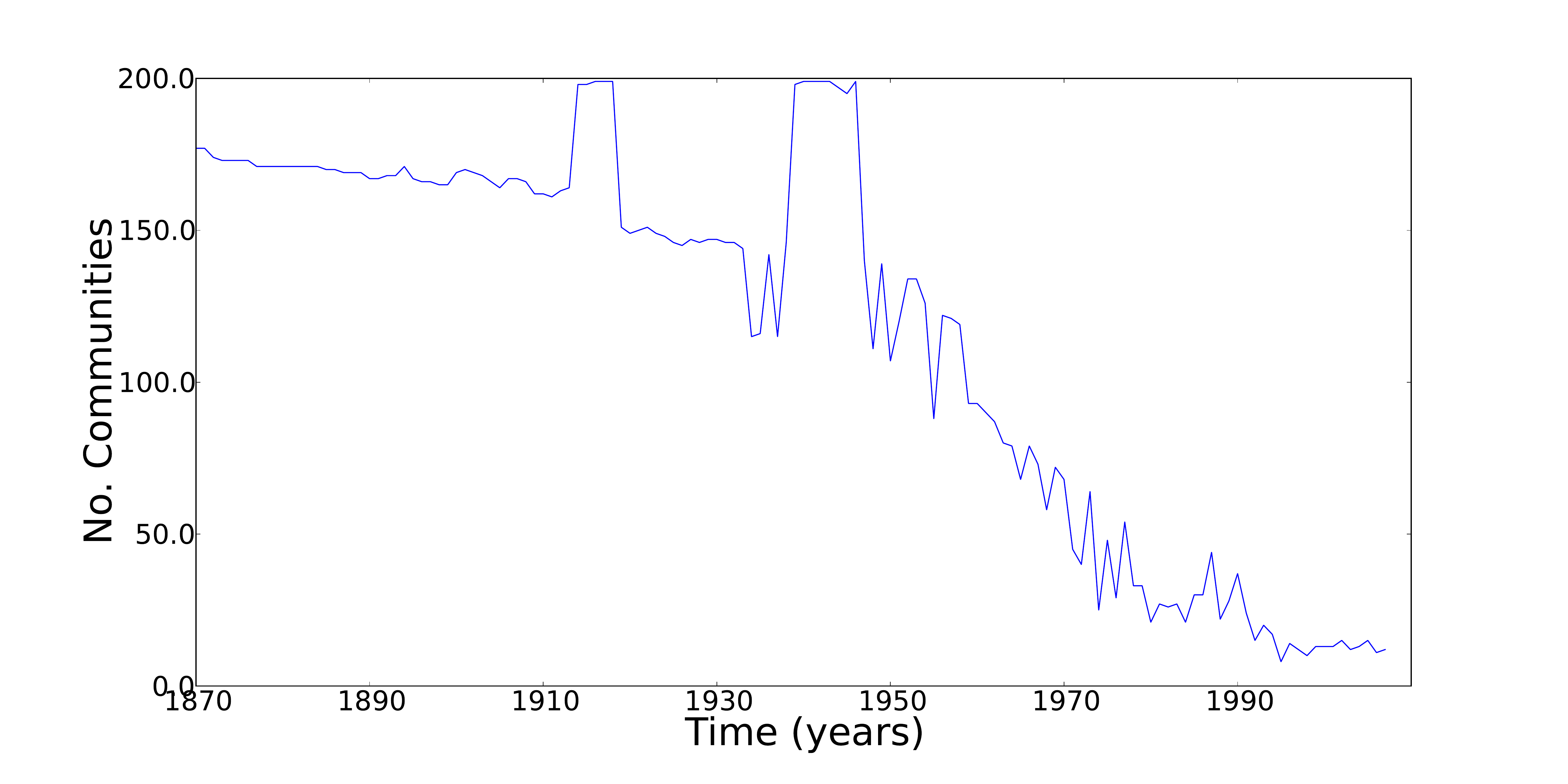}
\caption{Optimal stability results for $\tau=1$. }
\label{fig::cow_stability_partition_zoomin_trade}
\end{subfigure}
\begin{subfigure}[t]{0.48\textwidth}
\includegraphics[width=\textwidth]{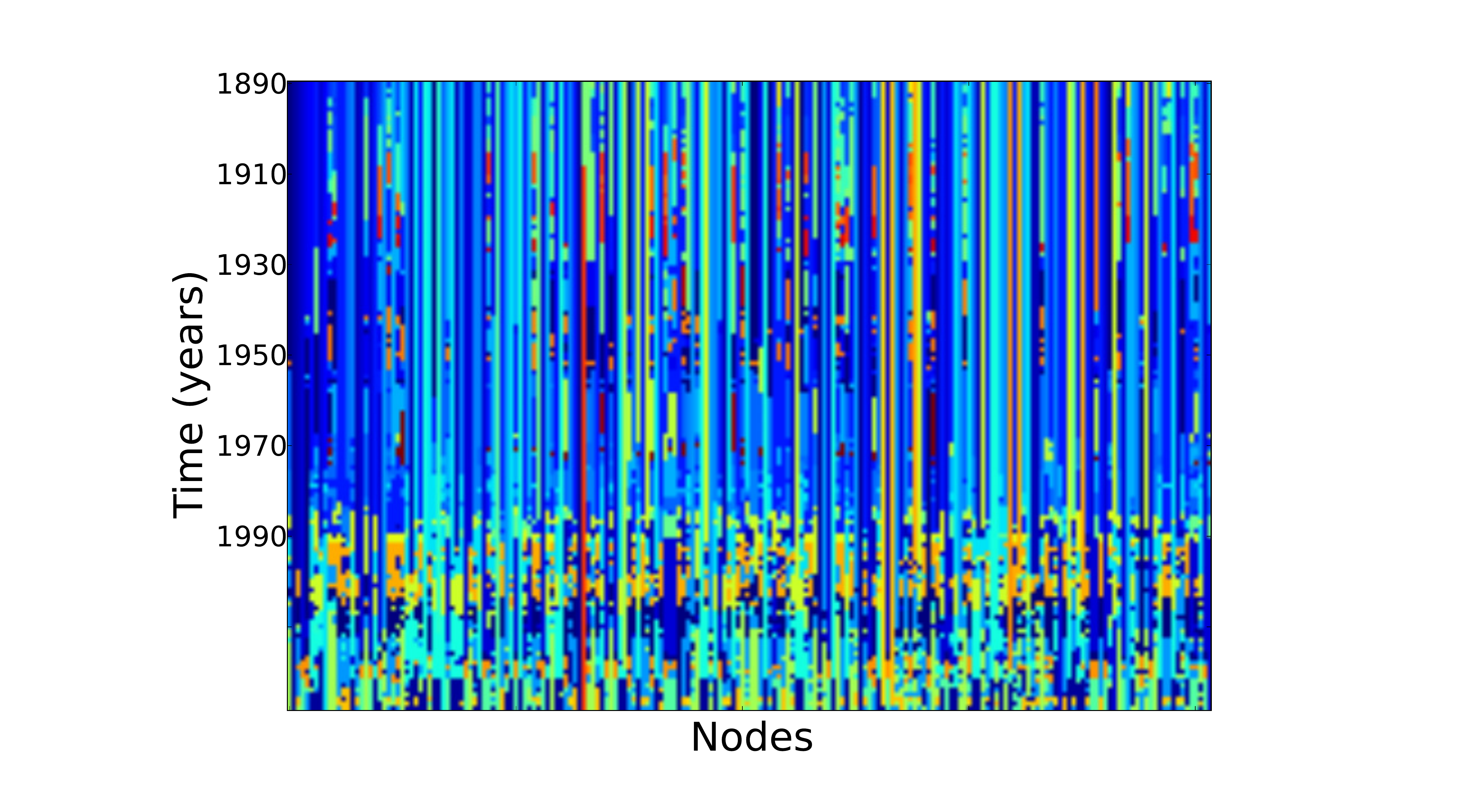}
\includegraphics[width=\textwidth]{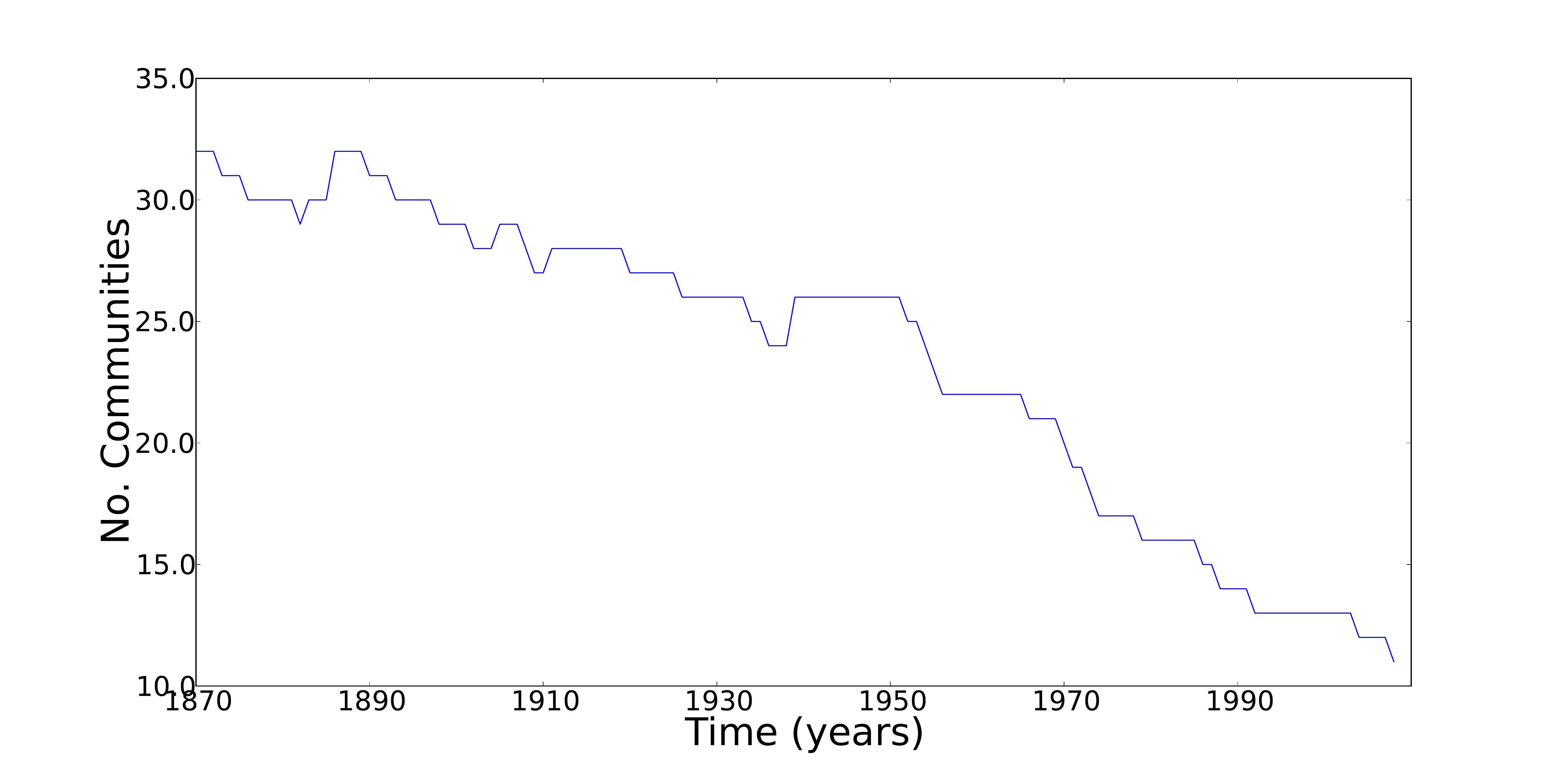}
\caption{Multislice results for $\omega=0.2$.}
\label{fig::cow_genlouvain_partition_0_2_trade}
\end{subfigure}
\begin{subfigure}[t]{0.48\textwidth}
\includegraphics[width=\textwidth]{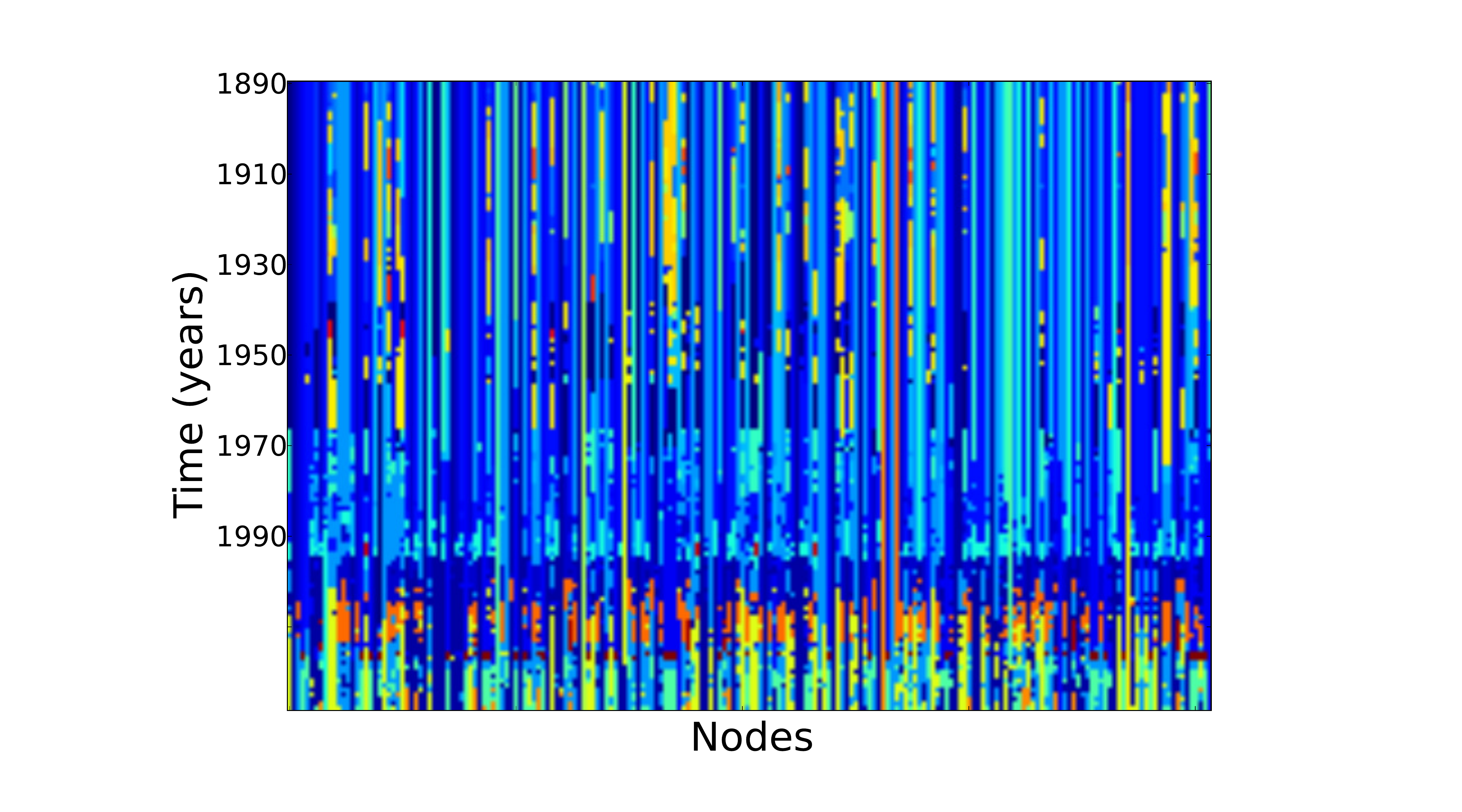}
\includegraphics[width=\textwidth]{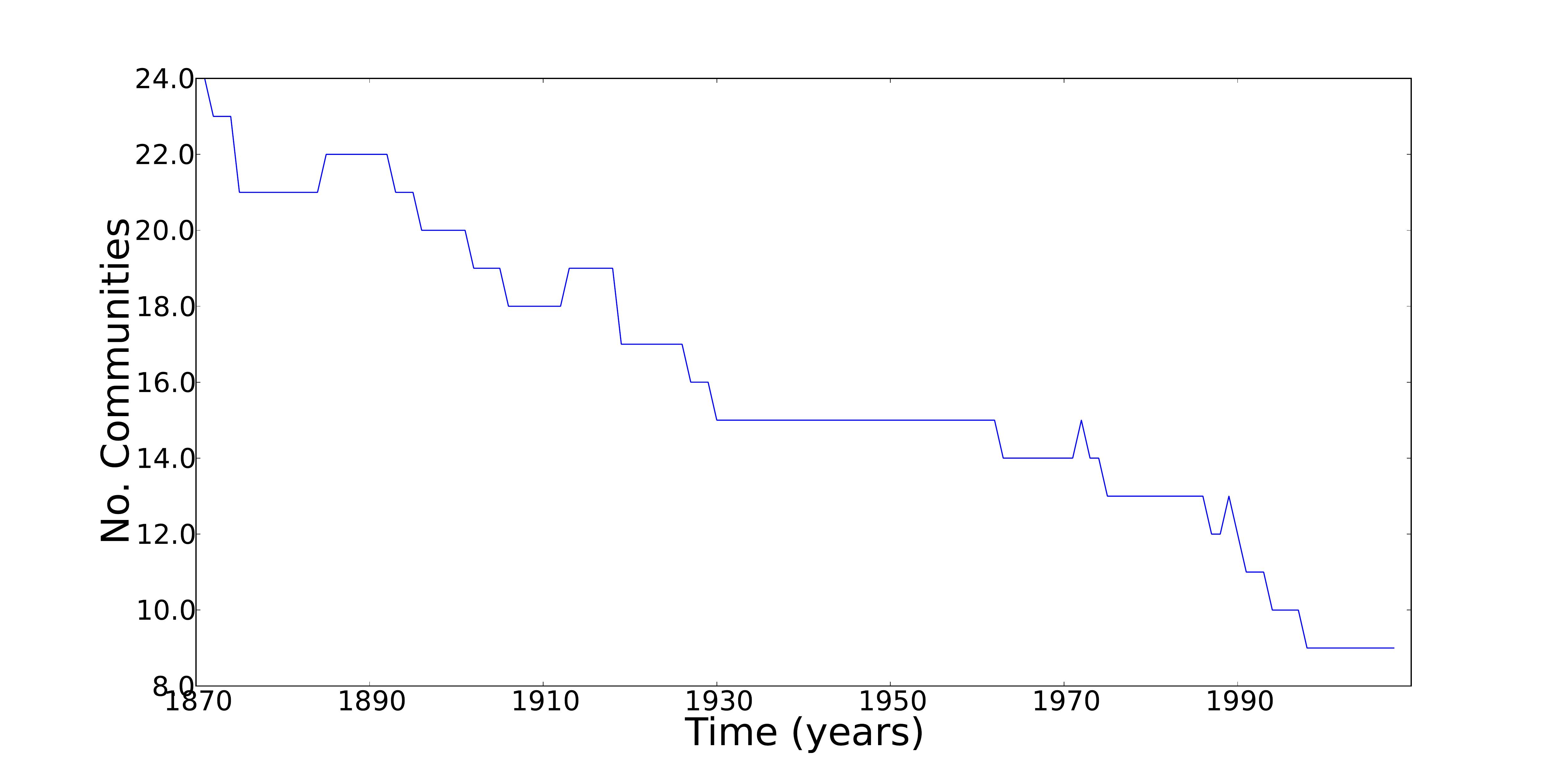}
\caption{Multislice results for $\omega=0.5$.}
\label{fig::cow_genlouvain_partition_0_5_trade}
\end{subfigure}
\begin{subfigure}[t]{0.48\textwidth}
\includegraphics[width=\textwidth]{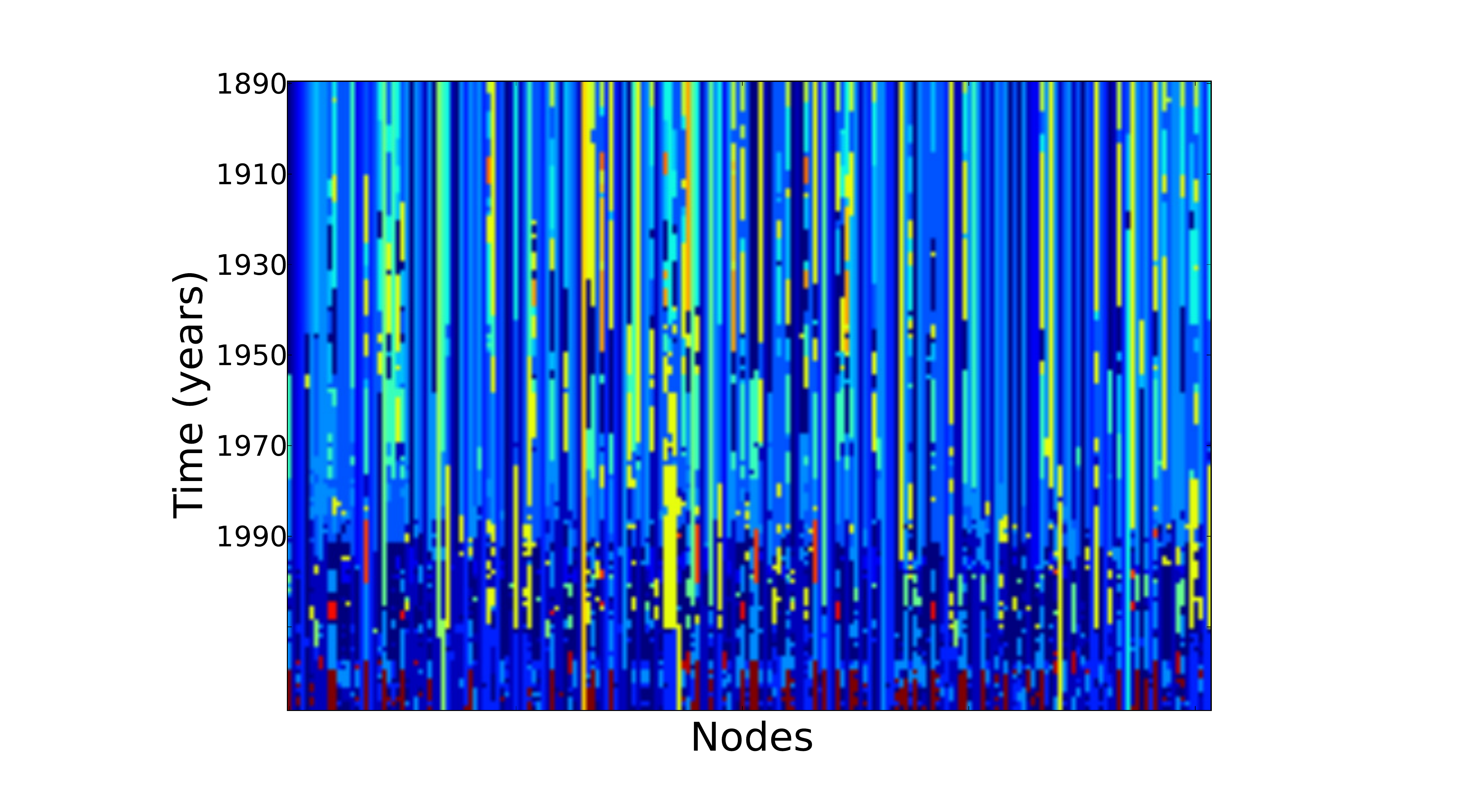}
\includegraphics[width=\textwidth]{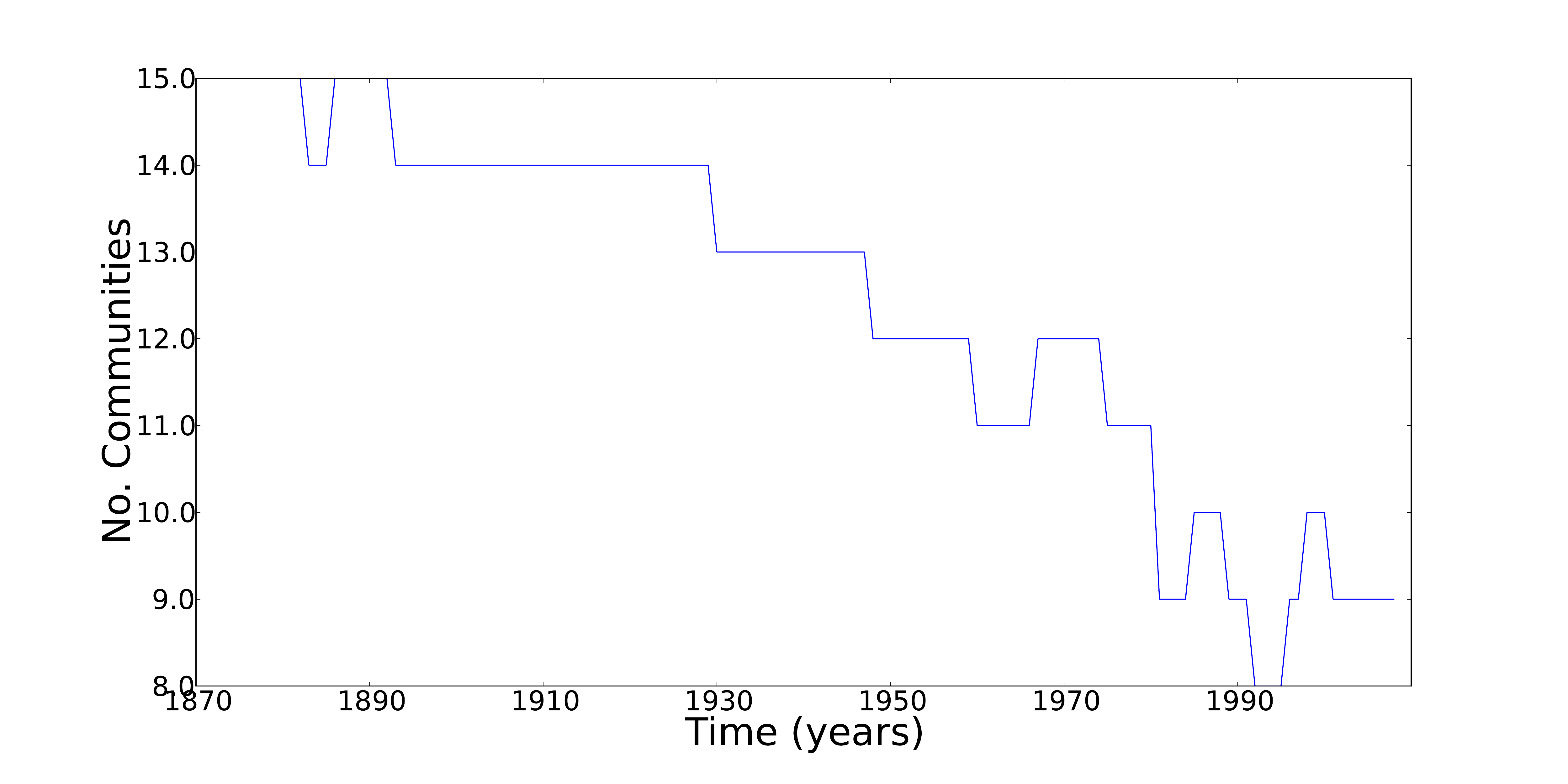}
\caption{Multislice results for $\omega=0.8$.}
\label{fig::cow_genlouvain_partition_0_8_trade}
\end{subfigure}

\caption{\textbf{Optimal stability and multislice modularity results for various $\omega$ for the international trade network.} Top plots represent community assignment through time, bottom plots show the number of communities at a given point in time.} \label{fig::cow_genlouvain_partition_zoomin}
\end{figure*}

\begin{figure*}
\centering
\includegraphics[width=0.5\textwidth]{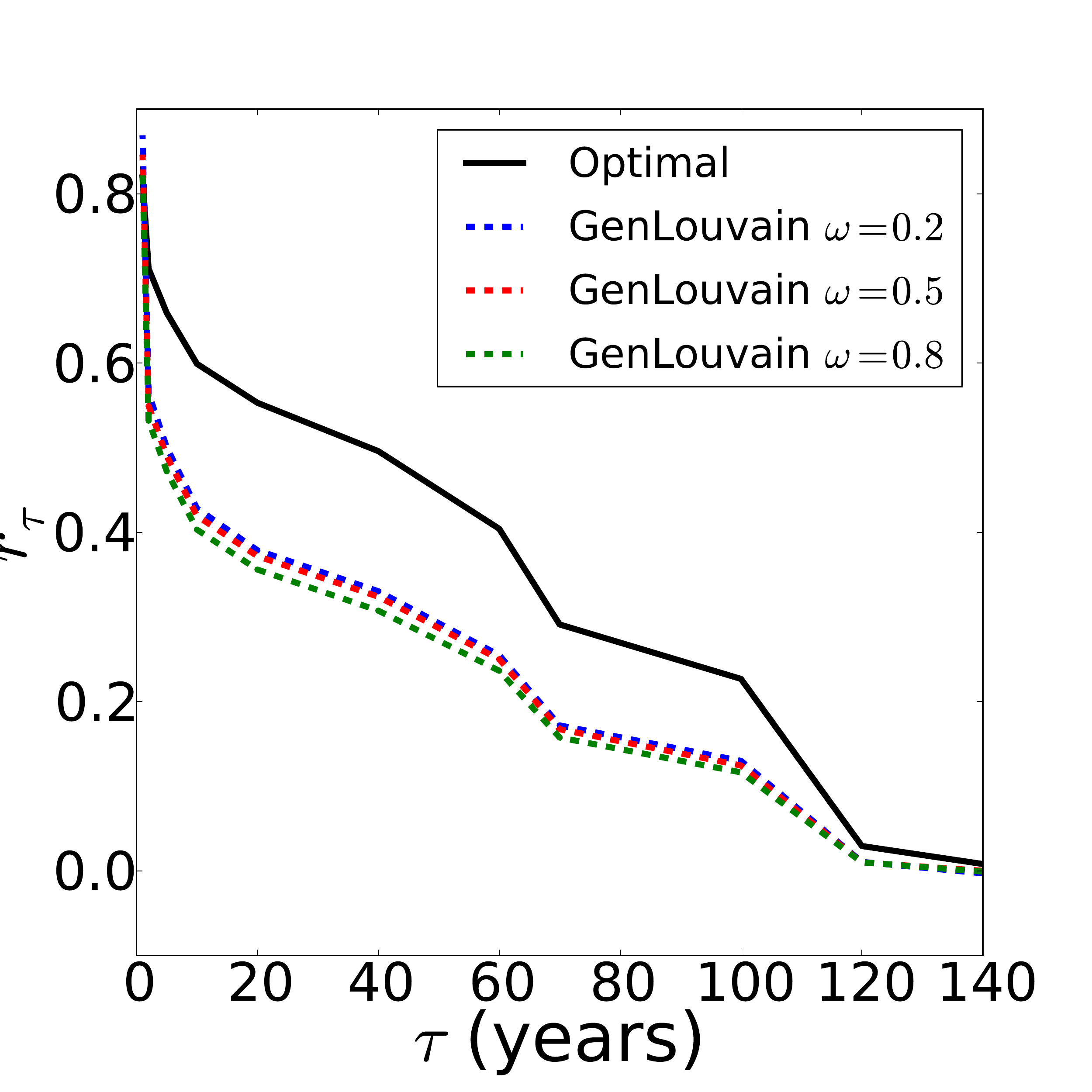}
\caption{\textbf{Comparison of stability values for multislice modularity for different $\omega$ and optimal partition for temporal stability for the International Trade Network.}} \label{fig::cow_comparison_genlouvain_stability}
\end{figure*}

As an example, we applied the GenLouvain algorithm on the three datasets considered in this paper. We used the code available at \url{http://netwiki.amath.unc.edu/GenLouvain/GenLouvain}. 
We calculated the best partitions according to the multislice modularity for various values of the parameter $\omega$ (which encodes the interlayer connection strength). 
The multislice modularity in general finds a small number of different clusters which change and evolve over time. 
The closest analog for the temporal stability case is the partition obtained for $\tau=1$, as this describes the temporal dynamics when considering only successive layers. 
We find marked differences between the stability values obtained by the multislice modularity partitions and the optimal stability ones (Fig. \ref{fig::Two_block_comparison_genlouvain_stability},  \ref{fig::sociopattern_comparison_genlouvain_stability} and \ref{fig::cow_comparison_genlouvain_stability}). 
In addition to this, it is easy to see from Figures \ref{fig::Two_blockcomparison}, \ref{fig::sociopattern_comparison} and \ref{fig::cow_genlouvain_partition_zoomin} that the results of the two methods differ significantly.
Multislice modularity produces larger and persistent communities over time due to the extra structure that it introduces in the network by the interlayer couplings, which effectively mixes time and space in the network in an unfolded static topology. 
The coupling parameter therefore stiffens the community structure by imposing (arbitrarily) stronger connection between time slices, resulting in a smaller number of communities with increasing $\omega$
The results for optimal stability at $\tau=1$ highlight on the contrary changes in the network as seen from a walker co-evolving with the network. 
This can be seen very clearly in the case of the two-block model and of the International Trade Network. 
In the former, multislice modularity yields few (3-4) communities which do not change over time significantly, while the temporal stability at $\tau=1$ yields a temporal partition that picks up with remarkable accuracy the cycles of the model. 
In the latter, the sparsity of links during the early years of the International Trade Network yields partitions mostly constituted by single-node or very small communities. 
%Coherently, we see in both methods that the number of communities found diminishes along with the increase in density of the yearly trade networks \ref{fig::stability_partition_zoomin_trade},\ref{fig::genlouvain_partition_0_2_trade}}, \ref{fig::genlouvain_partition_0_5_trade} and \ref{fig::genlouvain_partition_0_8_trade}. 
While the progressive integration of the trade network is evident in both methods (a decreasing number of communities with time (Fig. \ref{fig::cow_genlouvain_partition_0_2_trade}, \ref{fig::cow_genlouvain_partition_0_5_trade} and \ref{fig::cow_genlouvain_partition_0_8_trade}), optimal temporal stability for small $\tau$ is able to pick up distinctive, rapid changes in the network structure (e.g. the two World Wars), as opposed to multislice modularity.
So even when only considering as a naive heuristic as the number of communities, the advantage in uncovering rapid changes in the temporal structure that temporal stability has over the multislice modularity is evident . \\

In the light of these observations, we can conclude that the two methods provide two different and complementary lenses to study a given system; one strongly focused on the static persistent topological structure of the network, the other on dynamics.
It is also worse reminded the reader that by changing $\tau$, one changes the time-scale at which one is probing the community structure evolution. For example, micro-communities best describe the two-block model for $\tau=1$ as the model is rather sparse over short time-scales. Then, by increasing $\tau$, one sees the existence of the two-blocks, together with the shocks of the mixing; this effect is shown in the main text in the variation of information graphs.

%The exact same blabal, showing it is not an artifact due to the dataset used, but a genuine different between the two methods.
%This is supported also by the results on the Sociopatterns school data. 
%Here multislice modularity identifies as the core structure that of the different classes, which is already evident at the level of the static aggregate network, while the optimal partition for $\tau=1$ reveals a host of micro-temporal patterns, as for example the lunch and class breaks. 
%This effect makes sense since the 

%\begin{figure*}
%\centering
%\includegraphics[width=0.8\textwidth]{plots/COW_Trade_3.0/optimal_stability_tau_1_partition_across_snapshots.pdf}
%\includegraphics[width=0.8\textwidth]{plots/COW_Trade_3.0/optimal_stability_tau_1_number_of_communities_across_snapshots.pdf}
%\caption{Results for optimal partition for temporal stability with $\tau=1$ for the International Trade Network data.} \label{fig::stability_partition_zoomin}
%\end{figure*}

%\bibliographystyle{apsrev}
%\bibliography{complete_bib}

\end{document}